\documentclass[aps,twocolumn,reprint, superscriptaddress]{revtex4-2}

\usepackage{amsmath,amssymb}
\usepackage[latin9]{inputenc}

\usepackage{physics}

\usepackage[color= blue!20]{todonotes}
\usepackage[normalem]{ulem}
\usepackage{color}
\usepackage{mathtools}

\usepackage{graphicx}
\usepackage{hyperref}

\usepackage{xfrac}

\renewcommand{\bra}[1]{\ensuremath{\langle{#1}|}}
\renewcommand{\ket}[1]{\ensuremath{|{#1}\rangle}}

\newcommand{\dbs}{\delta_\mr{BS}}
\newcommand\varpm{\mathbin{\vcenter{\hbox{
            \oalign{\hfil$\scriptscriptstyle+$\hfil\cr
                \noalign{\kern-.5ex}
                $\scriptscriptstyle(\hspace{-0.1ex}{-}\hspace{-0.1ex})$\cr}
}}}}

\newcommand{\D}{\Delta}

\newcommand\mr{\mathrm}

\begin{document}
\bibliographystyle{naturemag}

\title{Engineering phonon-phonon interactions in multimode circuit quantum acousto-dynamics}

\author{Uwe von L\"upke}
\email[]{vluepkeu@ethz.ch}
\affiliation{Department of Physics, ETH Z\"{u}rich, 8093 Zurich, Switzerland}
\affiliation{Quantum Center, ETH Z\"{u}rich, 8093 Z\"{u}rich, Switzerland}
\author{Ines C. Rodrigues}
\affiliation{Department of Physics, ETH Z\"{u}rich, 8093 Zurich, Switzerland}
\affiliation{Quantum Center, ETH Z\"{u}rich, 8093 Z\"{u}rich, Switzerland}
\author{Yu Yang}
\affiliation{Department of Physics, ETH Z\"{u}rich, 8093 Zurich, Switzerland}
\affiliation{Quantum Center, ETH Z\"{u}rich, 8093 Z\"{u}rich, Switzerland}
\author{Matteo Fadel}
\affiliation{Department of Physics, ETH Z\"{u}rich, 8093 Zurich, Switzerland}
\affiliation{Quantum Center, ETH Z\"{u}rich, 8093 Z\"{u}rich, Switzerland}
\author{Yiwen Chu}
\email[]{yiwen.chu@ethz.ch}
\affiliation{Department of Physics, ETH Z\"{u}rich, 8093 Zurich, Switzerland}
\affiliation{Quantum Center, ETH Z\"{u}rich, 8093 Z\"{u}rich, Switzerland}

\maketitle
\textbf{In recent years, remarkable progress has been made towards encoding and processing quantum information in the large Hilbert space of bosonic modes.}
\textbf{Mechanical resonators are of great interest for this purpose, since they confine many high quality factor modes into a small volume and can be easily integrated with many different quantum systems.}
\textbf{An important yet challenging task is to create direct interactions between different mechanical modes. }
\textbf{Here we demonstrate an in-situ tunable beam-splitter-type interaction between several mechanical modes of a high-overtone bulk acoustic wave resonator.}
\textbf{The engineered interaction is mediated by a parametrically driven superconducting transmon qubit, and we show that it can be tailored to couple pairs or triplets of phononic modes.}
\textbf{Furthermore, we use this interaction to demonstrate the Hong-Ou-Mandel effect between phonons. 
Our results lay the foundations for using phononic systems as quantum memories and platforms for quantum simulations.}\\

Mechanical degrees of freedom are a particularly interesting platform for quantum information processing, as they typically have long coherence times and can be combined with many other quantum systems \cite{Chu20}. Circuit quantum acoustodynamics (cQAD) systems, where a superconducting qubit is coupled to GHz frequency acoustic modes, have recently been engineered \cite{Chu17, Satzinger18, Arrangoiz2019} and used to demonstrate the generation and measurement of nontrivial quantum states \cite{vonLupke22, bild2022schr, Chu18, Sletten19, Arrangoiz2019} and entanglement between mechanical modes \cite{wollack2022quantum}.
Furthermore, the small mode volumes, low-crosstalk, and high coherence times of acoustic modes have made cQAD devices the target platform of a recent proposal for the realization of a quantum random access memory (QRAM) \cite{Hann2019}. \\

The engineering of a hardware efficient QRAM would represent a major step towards the realization of a scalable quantum computing architecture \cite{Hann2019, Pechal2019, chamberland2022building}. 
In the specific case of cQAD, the QRAM would rely on a network of phononic modes to act as a quantum memory and could be implemented compactly in a single chip. 
So far, the main challenge associated with the physical implementation of a QRAM in cQAD is the generation of a phononic SWAP gate, i.e. an operation that allows for a direct exchange of quanta between mechanical modes. 
This can be engineered via a beam-splitter interaction, a coupling mechanism which has already been studied between photonic modes \cite{Gao18, Rodrigues21}, in optomechanical systems \cite{Aspelmeyer14}, trapped ions \cite{Toyoda15}, and between mechanical resonators in the classical regime \cite{Pino22, halg2022strong}. 
When brought to the quantum regime, this phononic SWAP gate will not only become a building block of QRAMs, but will also offer exciting possibilities for quantum metrology \cite{munro2002weak} and simulation \cite{wang2020efficient, Huh15, Sparrow18}, bosonic encodings \cite{teoh2022dual, Lau16}, and the study of quantum mechanical interference phenomena between phonons, such as non-reciprocal phononic control if extended to more than two modes \cite{Pino22}.\\

In this work, we demonstrate a beam-splitter interaction between multiple phonon modes of a high overtone bulk acoustic wave resonator (HBAR) coupled to a superconducting transmon qubit. 
We create this interaction by applying two off-resonant drives on the qubit \cite{zhang2019engineering} such that it acts as a nonlinear mixing element. 
We first study the effects of this bichromatic driving through qubit spectroscopy, observe the generation of multiple sidebands, and show how these sidebands mediate the desired beam-splitter coupling. 
Having realized this interaction, we then perform time domain experiments to demonstrate SWAP and $\sqrt{i\mr{SWAP}}$ gates, subsequently using the latter to demonstrate entanglement between two acoustic overtone modes of our HBAR. 
Furthermore, by choosing another parameter regime, we create an interference between three phononic modes and explore the multi-mode dynamics governing the system.
Finally, we utilize the beam-splitter interaction to exchange multiple excitations between modes and observe Hong-Ou-Mandel interference \cite{Gao18, Toyoda15, Kobayashi16, Qiao23, lopes2015atomic} between macroscopic mechanical modes.\\
 
The device used in this work is a cQAD system where a superconducting qubit is flip-chip bonded to a high overtone bulk acoustic wave resonator \cite{Chu18}.  
The qubit is a 3D transmon with a frequency of $\omega_q =  2\pi\cdot 5.97\,$GHz, an energy relaxation time of $T_{1} = 9.5\,\mu$s, a Ramsey decoherence time of $T_{2}^{*} = 7.2\,\mu$s, and an anharmonicity $\alpha = 2\pi\cdot218\,$MHz. 
The longitudinal free spectral range (FSR) of the HBAR is approximately $2\pi\cdot 12.63\,$MHz, and the two subsystems are coupled through a piezoelectric transducer that mediates a Jaynes-Cummings (JC) interaction with a coupling strength of $g_m=2\pi\cdot257\,$kHz.
The device is housed in a 3D aluminum cavity which we use to both shield the qubit from its environment and read out its state via the dispersive interaction between qubit and cavity. \\

While this cQAD system has been studied in previous works in both the dispersive \cite{vonLupke22} and the resonant coupling regimes \cite{bild2022schr}, here we focus on direct phonon-phonon interactions that arise when two parametric drives are applied to the qubit. The Hamiltonian of our system in the presence of these drives is given by
\begin{eqnarray}
    H&=&\omega_q q^\dag q -\frac{\alpha}{2}{q^\dag}^2 q^2 \nonumber \\ 
    &+& \sum_m \left[\omega_m m^\dag m + g_m (m^\dag q + m q^\dag)\right] +H_\textrm{qd},~~~~~\label{eq:bareHamiltonian}
\end{eqnarray}
where we assume $g_m$ to be real. Here the first two terms describe the qubit as an anharmonic mode with lowering operator $q$. The sum over phonon modes $m=a,b,c,...$ with frequencies $\omega_m$ and lowering operators $m$ includes their energies as well as their JC interaction with the qubit. The last term, given by $H_\mr{qd}=\left( \Omega_{1} e^{- i \omega_{1} t} + \Omega_{2} e^{- i \omega_{2} t} \right){{q}^\dag} + \mathrm{h.c.}$, describes two off-resonant microwave drives applied to the qubit with frequencies $\omega_{1,2}$. 
As shown in previous works \cite{gao2019,zhang2019engineering,Hann2019}, the drives, together with two phonon modes $a$ and $b$, can participate in a four-wave mixing process mediated by the Josephson non-linearity of the superconducting qubit. In particular, when the resonance condition $\Delta_{21} \equiv \omega_2-\omega_1 = \omega_b-\omega_a$ is satisfied, Eq.~(\ref{eq:bareHamiltonian}) leads to a bilinear coupling between the phonon modes. Even though this picture is quantitatively accurate for large phonon-phonon detunings and small drive strengths, we now present a framework that extends this picture to address the case of large drive strengths and small phonon-phonon detunings. Furthermore, our analysis readily lends itself to systems with many bosonic modes by explicitly considering processes involving multiple drive photons. \\
\begin{figure}[tbp]
    \centerline{\includegraphics[trim = {0cm, 0.5cm, 0.0cm, 0.0cm}, clip=True,scale=0.51]{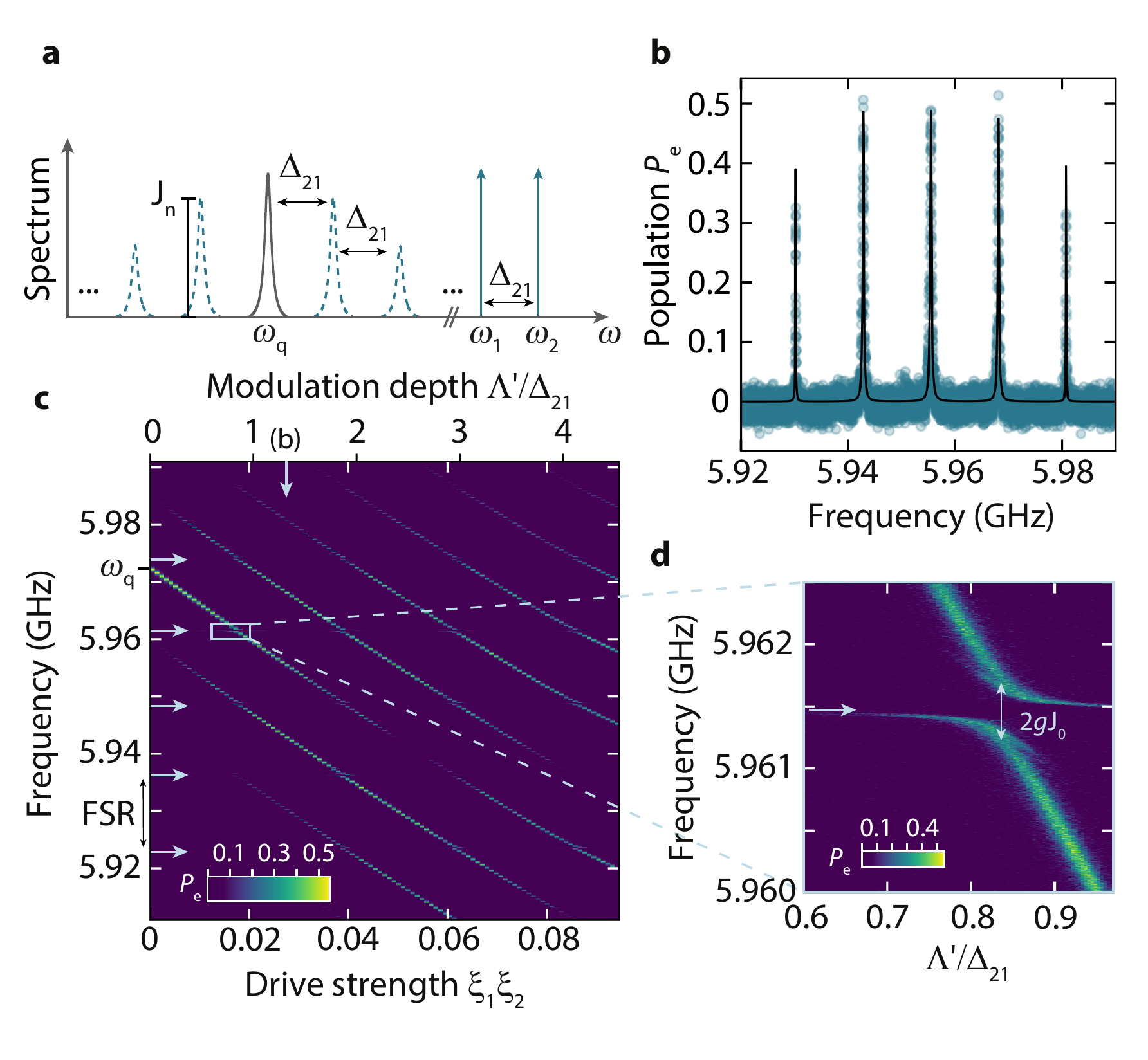}}
    \caption{\textsf{\textbf{Effects of bichromatic driving on a transmon qubit coupled to an HBAR.} \textbf{a}. Schematic illustration of the spectrum of a bichromatically driven qubit. The blue vertical lines represent the two drives, the black Lorentzian peak represents the qubit resonance, and the dashed Lorentzian peaks in blue represent the generated sidebands with amplitudes $J_n$ for the $n$th sideband. \textbf{b}. Qubit population $P_\textrm{e}$ during spectroscopy for a drive strength of $\xi_{1}\xi_{2} \sim 0.0274$. Circles are data and the black line is a theoretical curve (see Supplementary Information \cite{SI}) for the measured qubit population $P_\textrm{e}$ when sweeping a probe tone over the qubit sidebands. \textbf{c}. Qubit spectroscopy for different values of $\xi_{1}\xi_{2}$. Top x-axis indicates the corresponding modulation depth $\Lambda '/\Delta_{21}$. The vertical arrow indicates the linescan shown in \textbf{b} and the horizontal arrows indicate the phonon mode frequencies of the HBAR. \textbf{d}. Zoom-in of one of the qubit-phonon anti-crossings in \textbf{c}.}}
    \label{fig:Fig1}
\end{figure}

We first consider only the effect of the drives on the qubit itself. Due to the transmon anharmonicity, going into the displaced frame of the drives results in a modulated AC Stark shift of the qubit frequency, given by \cite{SI} 
\begin{equation}
    H_\mr{Stark} = \left[- 2 \alpha (\xi_{1}^{2} + \xi_{2}^{2}) - 4\alpha \xi_{1} \xi_{2} \cos{(\Delta_{21}t)} \right]{{q}^\dag} {q}, \label{eq:stark}
\end{equation} 
with the dimensionless drive strengths  $\xi_{j}=\Omega_{j}/\Delta_{j}$, where $\Delta_{j}=\omega_{j}-\omega_q$ for $j\in\{1,2\}$. This shift has both a time-independent and a time-dependent contribution, the latter arising from the beating between the two drives, which modulates the qubit frequency with $\Delta_{21}$. 
As usual for a frequency modulated system \cite{Strand13,Naik2017,kervinen2019landau} and as illustrated in Fig.~\ref{fig:Fig1}\textbf{a}, this gives rise to the appearance of multiple qubit sidebands separated by $\Delta_{21}$, whose amplitudes are given by $J_n \left( \frac{\Lambda}{\Delta_{21}} \right)$. Here $J_n(x)$ is the Bessel function of the first kind for a given sideband number $n$, and $\Lambda = -4\alpha \xi_{1} \xi_{2}$. We note that, due to the interplay of the parametric drives with the third energy level of the qubit, $H_\mr{Stark}$ acquires a correction, which we derive using time-independent perturbation theory (see Supplementary Information  \cite{SI}). In the following we use the corrected value for the modulation depth which we label $\Lambda'$. 
Furthermore, we will use the shorthand $J_n=J_n \left( \frac{\Lambda'}{\Delta_{21}} \right)$. \\

We confirm these effects experimentally via two-tone spectroscopy. Specifically, we sweep a weak probe signal across the qubit frequency while the off-resonant drives are turned on and subsequently measure the resulting qubit population using dispersive readout. 
As expected, we find multiple resonances separated by $\Delta_{21}$ with different peak heights, which are the qubit sidebands described above (cf. Fig.~\ref{fig:Fig1}\textbf{b}).
The measured steady-state population of the qubit is quantitatively described in the same way as in a regular qubit spectroscopy experiment \cite{schuster2005ac} with the probe strength adjusted by the sideband amplitude \cite{SI}, as shown by the continuous black line in Fig.~\ref{fig:Fig1}\textbf{b}. 
After repeating the measurement for a range of parametric drive strengths $\xi_{1} \xi_{2}$ (with $\xi_1=\xi_2$) we find the result shown in Fig.~\ref{fig:Fig1}\textbf{c}, where we observe multiple diagonal lines spaced in frequency by $\Delta_{21}$ and with varying intensity. 
These qubit sidebands shift to lower frequencies with increasing drive power, as expected from the Stark shift described by the first term in Eq.~(\ref{eq:stark}).\\

The JC interaction between the driven qubit and the phonon modes results in anti-crossings where the frequency of a sideband matches that of a phonon mode, as shown in Fig.~\ref{fig:Fig1}\textbf{c} and \textbf{d}. 
However, the effective qubit-phonon coupling strength is scaled by the amplitude of the sideband closest to the phonon mode. Therefore, the gap of the anti-crossing will be reduced from $2g_m$ to $2J_ng_m$, as indicated for $n=0$ in Fig.~\ref{fig:Fig1}\textbf{d}.\\

In the dispersive regime, where all the qubit sidebands and phonon modes are far detuned, it is useful to enter the interaction picture of the sideband-mediated qubit-phonon coupling via the Schrieffer-Wolff transformation \cite{schrieffer1966relation}. After applying the rotating wave approximation (RWA), we can identify two effects in the resulting effective Hamiltonian. First, there is a frequency shift of the phonon modes, due to their hybridization with the qubit \cite{Gely2021}, such that the phonon frequency in the presence of the driven qubit is $\omega_m+\delta_m$ with    
\begin{equation}
    \delta_m = g_m^2\sum_{n} \frac{ J_n^2 }{\tilde{\Delta}_{m}  - n\Delta_{21}}
    \label{eq:deltaBS_main}~,  
\end{equation}
where $\tilde{\Delta}_{m}=\omega_m - \tilde{\omega}_{q}$ is the detuning between phonon mode $m$ and the Stark shifted qubit. 
We see that that a phonon mode's frequency shift is dominated by the sideband for which the denominator in Eq.~(\ref{eq:deltaBS_main}) is smallest \cite{SI}. Second, while the Schrieffer-Wolff transformation typically eliminates the JC coupling term between qubit and phonons, in our case it also gives rise to phonon-phonon coupling terms. For example, the coupling between two neighboring phonon modes $b$ and $c$ is given by $g_\mr{bc}(b^\dag c + b c^\dag)$, with 
\begin{equation}
    g_\mr{bc} = g_b g_c \sum_{n} \frac{ J_n J_{n+1} }{\tilde{\Delta}_{b}  - n\Delta_{21}} 
    \label{eq:gBS_main}~, 
\end{equation}
when $\Delta_{21}=\omega_c - \omega_b + \delta_c - \delta_b$, such that this term remains after the RWA. 
Similarly, next-nearest-neighboring phonon modes $a$ and $c$ experience a coupling of $g_\mr{ac}(a^\dag c + a c^\dag)$, with 
\begin{equation}
    g_\mr{ac} = g_a g_c \sum_{n} \frac{ J_n J_{n+2} }{\tilde{\Delta}_{a}  - n\Delta_{21}} 
    \label{eq:gac_main}~,
\end{equation}
when $2\Delta_{21}=\omega_c - \omega_a + \delta_c - \delta_a$.\\

The numerator of Eq.~(\ref{eq:gBS_main}), which contains the product of two successive Bessel functions, represents the physical process of the qubit converting one photon between the parametric drives. 
The frequency conversion of the drive photons compensates for the energy difference between the phonon modes, making the beam-splitter interaction resonant. 
Interestingly, the effective coupling strength for this process does not become larger monotonically with increasing drive strengths $\xi_1\xi_2$. 
Instead, the speed of the single photon conversion is reduced in favor of multi-photon processes, for example converting two drive photons to bridge the energy gap between phonon modes with a frequency difference of $2\D_{21}$, as shown in Eq.~(\ref{eq:gac_main}). 
A more detailed derivation of the different transformations  and their effects on the system Hamiltonian can be found in the Supplementary Information \cite{SI}. \\

\begin{figure*}[tbp]
    \centerline{\includegraphics[trim = {1cm, 0.3cm, 1cm, 0cm}, clip=True, width=0.99\textwidth]{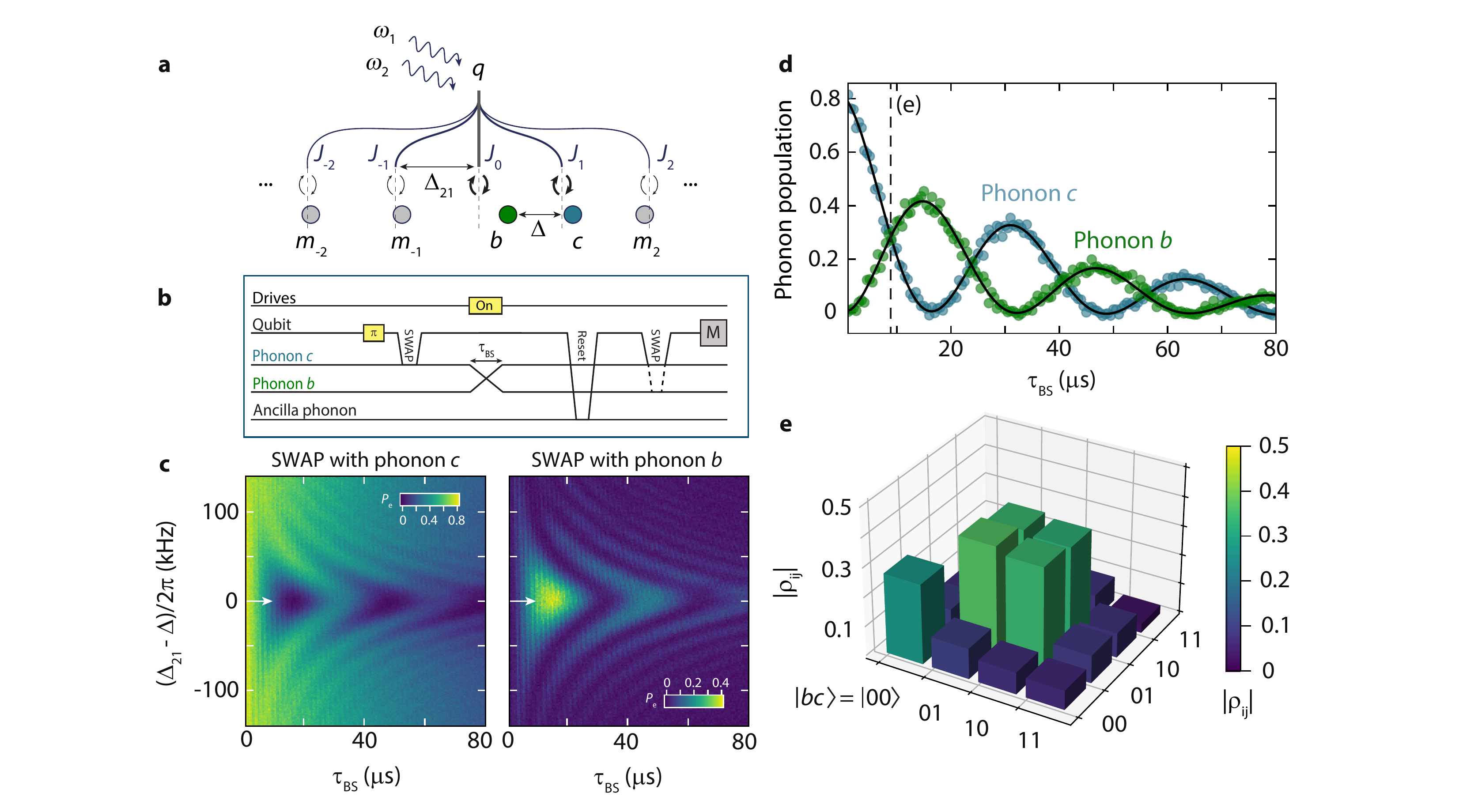}}
    \caption{\textsf{\textbf{Beam-splitter interaction between two acoustic modes.} \textbf{a}. Schematic representation of the beam-splitter coupling between two mechanical modes $b$ and $c$ mediated by the qubit sidebands. The frequency difference between the drives is given by $\Delta_{21}$, whereas the acquired unique spacing between the two neighboring modes of interest is given by $\Delta$. \textbf{b}. Pulse sequence used in the experiment. \textbf{c}. Phonon population versus detuning ($\Delta_{21} - \Delta$) and interaction time $\tau_\textrm{BS}$. We perform the pulse sequence described in \textbf{b} while changing the drive frequency $\omega_2$ and read out the population in either mode $c$ (left plot) or mode $b$ (right plot). The horizontal white arrow indicates the curves shown in \textbf{d}. \textbf{d}. Rabi oscillations between the two mechanical modes when $\Delta_{21} = \Delta$. The vertical dashed line shows the interaction time $\tau_\textrm{BS} = 8.0\,\mu$s at which the tomography experiment in \textbf{e} was performed. Black lines are fits to a decaying sinusoidal function.  \textbf{e}. Reconstructed density matrix for joint phonon state after a 50:50 beam splitter interaction. Both the colors and height indicate the magnitude of the matrix elements.}}
    \label{fig:Fig2}
\end{figure*}

The dependence of the qubit sidebands on the Bessel functions is what allows us to choose different combinations of coupling strengths between phonon modes and frequency shifts throughout this work. Naively, it might seem that, due to the equal frequency spacing of the phonon modes, one cannot choose interactions between only a subset to be resonant. 
However, this is not the case. For instance, by choosing an appropriate modulation depth $\Lambda'/\Delta_{21}$, we can choose the amplitude of $J_0$ to be larger than those of the neighbouring sidebands, $J_1$ and $J_{-1}$. According to Eq.~(\ref{eq:deltaBS_main}), the phonon mode closest to the $0$th sideband will shift by a larger amount $(\propto J_0^2)$ than the adjacent phonon modes $(\propto J_1^2,~J_{-1}^2)$, giving rise to a unique frequency spacing between two phonons modes equal to $\Delta_{21}$ and promoting a beam-splitter interaction between them (cf. Fig.~\ref{fig:Fig2}\textbf{a}). 
If, on the other hand, we choose a regime where $J_0=J_1=-J_{-1}$, the three phonon modes $a$, $b$, and $c$ adjacent in frequency to the $n=-1,0, 1$ sidebands will be shifted equally, promoting beam-splitter interactions between these three modes. Note that in the latter case, the next-nearest-neighbor modes $a$ and $c$ are coupled via a two-photon conversion described by Eq.~(\ref{eq:gac_main}).\\

We now experimentally investigate the first case of coupling between only the two modes  $b$ and $c$ (cf. Fig.~\ref{fig:Fig2}\textbf{a}).
We do this by setting the modulation depth to $\Lambda'/\D_{21}=0.61$ such that $J_0=0.91$ and $J_1=0.29$. 
Our experimental protocol starts with swapping an excitation from the qubit into mode $c$ using the resonant JC interaction. Note that we use a third microwave drive, far detuned from the parametric drives, to independently adjust the frequency of the qubit for this swap operation and to compensate the Stark shift of the qubit from the parametric drives during the beam-splitter interaction to set $\tilde{\D}_b=2\pi\cdot 1.0\,$MHz. We then turn on the parametric drives for a variable time $\tau_\mr{BS}$ (cf. Fig.~\ref{fig:Fig2}\textbf{b}).
Afterwards, the qubit has a finite excited state population due to the off-resonant drives. We reset the qubit to its ground state by swapping its residual population to an ancillary phonon mode detuned by several FSRs from the modes of interest \cite{bild2022schr}. Finally, we swap the excitation from mode $b$ or $c$ into the qubit and measure its excited state population. \\

Repeating this experiment for different values of $\Delta_{21}$, we observe the expected chevron pattern produced by a beam-splitter type interaction between two modes, as shown in Fig.~\ref{fig:Fig2}\textbf{c}. 
Here, we vary $\Delta_{21}$ by only about $\pm1\%$, such that we can treat the modulation depth as constant. 
When $\Delta_{21}$ matches the unique detuning between the two modes $\Delta$, we satisfy the resonance condition for the four-wave mixing process, and the exchange of quanta between the modes becomes most efficient. This occurs for a modulation frequency of $(\D_{21}-\mr{FSR}) = - 2\pi\cdot 44\,$kHz, which matches our prediction from Eq.~(\ref{eq:deltaBS_main}). 
We plot the phonon mode populations for $\Delta_{21} = \Delta$ in Fig.~\ref{fig:Fig2}\textbf{d} and fit them each to a decaying oscillation, yielding a beam-splitter coupling rate of $g_\mr{bc}=2\pi\cdot 15.6\,$kHz. 
Note that the contrast of the oscillation in phonon mode $b$ is slightly lower than that for phonon mode $c$. 
This is a result of different decay rates between the two phonon modes, as well as a small but finite leakage to the next phonon mode, $m_{-1}$ (cf. Fig.~\ref{fig:Fig2}\textbf{a}). 
The microscopic origin of the different decay rates for different HBAR modes is a subject of ongoing research \cite{agnetta2023}.  \\

\begin{figure*}[tbp]
    \centerline{\includegraphics[trim = {0.9cm, 0.7cm, 0.3cm, 0cm}, clip=True, width=1\textwidth]{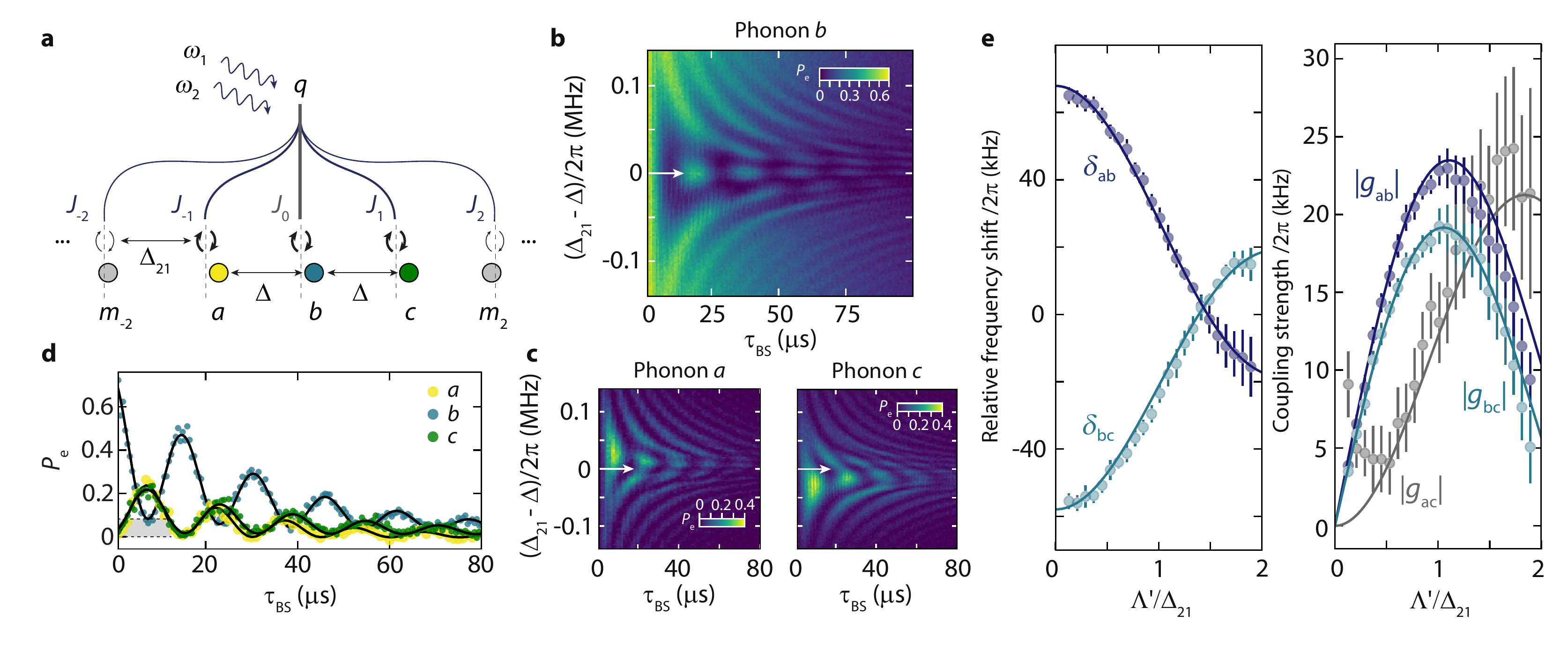}}
    \caption{\textsf{\textbf{Engineering a multimode coupling by tuning the parametric drive power.} \textbf{a} Schematic illustration of the beam-spliter coupling  between three modes. \textbf{b} Final phonon $b$ occupation versus detuning ($\Delta_{21} - \Delta$) and interaction time $\tau_\textrm{BS}$. \textbf{c} Final phonon $a$ ($c$) population versus detuning and interaction time. \textbf{d} Linescans of the individual phonon populations versus $\tau_\textrm{BS}$ for $\Delta_{21} = \Delta$, as indicated in \textbf{b} and \textbf{c} with horizontal white arrows. Black lines are fits to a decaying sinusoidal function and the grey shaded area points out the offset of the residual phonon $b$ occupation from zero.  \textbf{e} Relative frequency shifts and absolute coupling strengths between different phonon modes versus modulation depth $\Lambda '/\Delta_{21}$. The data (circles) was extracted from fitting data like the ones displayed in \textbf{b} and \textbf{c} for various values of $\Lambda '/\Delta_{21}$. The theory curves (full lines) are computed using Eq.~(\ref{eq:deltaBS_main}), Eq.~(\ref{eq:gBS_main}) and  Eq.~(\ref{eq:gac_main}). The error bars are extracted based on a $5\%$ induced change on the fitting residuals. For more details on the fitting routine and theory description of our multi-mode coupling as a three-level system, see Supplementary Information \cite{SI}.} }
    \label{fig:Fig3}
\end{figure*}
At the time $\tau_\mr{BS}=\pi/4g_\mr{BS}= 8.0\,\mu s$, indicated by a dashed line in Fig.~\ref{fig:Fig2}\textbf{d}, the interaction becomes a 50:50 beam-splitter or $\sqrt{i\mr{SWAP}}$ gate, which creates an entangled state between the two phonon modes.
We confirm this experimentally by performing two-qubit state tomography on the resulting state (cf. Fig.~\ref{fig:Fig2}\textbf{e}). 
Here, in contrast to the data shown in Figs.~\ref{fig:Fig2}\textbf{c} and \textbf{d}, we measure observables of both phonon modes in the same sequence, thereby accessing joint two-mode observables necessary for full state tomography.  
To quantify the entanglement created, we compute an overlap of the reconstructed density matrix with the maximally entangled state $\ket{bc} = (\ket{01}+e^{i\phi}\ket{10})/\sqrt{2}$ of $F_\mr{Bell}=0.69\pm0.01$, with $\phi$ chosen to optimize $F_\mr{Bell}$. 
This confirms the presence of entanglement between the two phonon modes. We attribute the difference between the reconstructed density matrix and the maximally entangled state to phonon decay during the $\sqrt{i\mr{SWAP}}$ gate and an imperfect state preparation of the initial Fock state in mode $c$. Details on the tomography procedure can be found in the Supplementary Information \cite{SI}.  \\ 

Having demonstrated a beam-splitter interaction between two phonon modes, we now move on to creating simultaneous interactions between three modes. 
To that end, we tune the modulation depth to $\Lambda'/\D_{21}=1.43$, such that $J_0=J_1=-J_{-1}=0.55$. 
In this regime, phonon modes $a$, $b$, and $c$ are shifted equally, such that $\Delta_{cb}=\D_{ba} \equiv \Delta$. This is schematically shown in Fig.~\ref{fig:Fig3}\textbf{a}. 
In this case, phonon mode pairs ($b,c$) and $(a,b)$ are coupled via Eq.~(\ref{eq:gBS_main}) while the mode pair ($a,c$) is coupled via Eq.~(\ref{eq:gac_main}), with $|g_{ab}| \approx|g_{bc}| \approx |g_{ac}|$. \\

In order to explore the dynamics of this three mode coupling scheme, we perform an experiment analogous to the one presented in Fig.~\ref{fig:Fig2}. 
Specifically, we load an excitation into phonon mode $b$, turn on the parametric drives, thereby activating beam-splitter interactions between all three modes, and finally measure their population.  
As before, we sweep the interaction time $\tau_\mr{BS}$ and the modulation frequency $\D_{21}$, with $\tilde{\D}_b=2\pi\cdot1.0\,$MHz. 
The results are shown in Figs.~\ref{fig:Fig3}\textbf{b} and \textbf{c}. While they show the expected qualitative aspects of the excitation swapping between all three modes, we observe two interesting features.
First, when $\D_{21}=\D$, the initial excitation in mode $b$ flows to modes $a$ and $c$ with approximately equal rates, as can be seen in Fig.~\ref{fig:Fig3}\textbf{d}
However, the excitation does not swap fully to modes $a$ and $c$, which is visible from the reduced oscillation contrast (see grey shaded area in Fig.~\ref{fig:Fig3}\textbf{d}). 
While counter-intuitive at first, this is the expected behavior of a three mode system with coupling between all mode pairs. 
The coupling between modes $a$ and $c$ hybridizes them into new normal modes with frequencies shifted by the coupling strength. As a result, the coupling between these normal modes and mode $b$ is no longer resonant, resulting in the reduced oscillation contrast we observe. We note that the frequency of the population exchange observed in Fig. \ref{fig:Fig3}\textbf{d} of $2\pi\cdot 64\,$kHz is in good agreement with theoretical calculations. \\

The second observation is that the data in Fig. \ref{fig:Fig3}\textbf{c} for mode $a$ is approximately the mirror image of mode $c$ with respect to $\D_{21}-\Delta = 0$. 
For instance, when $\D_{21}-\Delta> (<)~0$, the initial excitation in mode $b$ predominantly flows to mode $a$ ($c$). 
While the roles of modes $a$ and $c$ are symmetric when $\D_{21} = \Delta$, this symmetry is broken away from the resonance condition due to the coupling between modes $a$ and $c$ and the resulting normal mode splitting. 
A detailed explanation for both of these effects is presented in the Supplementary Information \cite{SI}. \\

While we present experimental details on two interesting values of modulation depth, we note that we can tune from one regime to the other by changing the drive powers, thereby observing a gradual change in both coupling strength and relative detuning as shown in Fig.~\ref{fig:Fig3}\textbf{e}. 
To acquire the effective interaction strengths between the three modes as well as their respective phonon frequency shifts, we perform the experiment shown in Fig.~\ref{fig:Fig3}\textbf{b} and \textbf{c} for different values of $\xi_1\xi_2$, thereby varying $\Lambda'/\D_{21}$. 
We then fit the measured phonon populations to a set of coupled equations of motion with beam-splitter couplings $g_{mk}$ and relative phonon detunings $\delta_{mk}$ as free parameters ($m,k\in\{a,b,c\}$). 
Details on the fitting procedure can be found in the Supplementary Information \cite{SI}. 
The fit results are plotted alongside Eqs.~(\ref{eq:deltaBS_main}), (\ref{eq:gBS_main}), and (\ref{eq:gac_main}) with no free parameters in Fig.~\ref{fig:Fig3}\textbf{e} and show good agreement between experiment and theory.
The observed difference between $|g_{ab}|$ and $|g_{bc}|$ is a result of the different relative contributions from the sidebands in Eq. (\ref{eq:gBS_main}) depending on the position of the phonon modes involved. 
Notably, the observed reduction of 
$|g_{ab}|$ and $|g_{bc}|$ for larger modulation depths, as well as the accompanying increase in 
$|g_{ac}|$ are well captured by the theory.
We emphasize that previous works have only investigated a much smaller range of modulation depths, so that these effects were not evident \cite{gao2019, zhang2019engineering, teoh2022dual, chapman2022high}. \\

\begin{figure}[tbp]
    \centerline{\includegraphics[trim = {0cm, 1cm, 0.0cm, 0.0cm}, clip=True,scale=0.51]{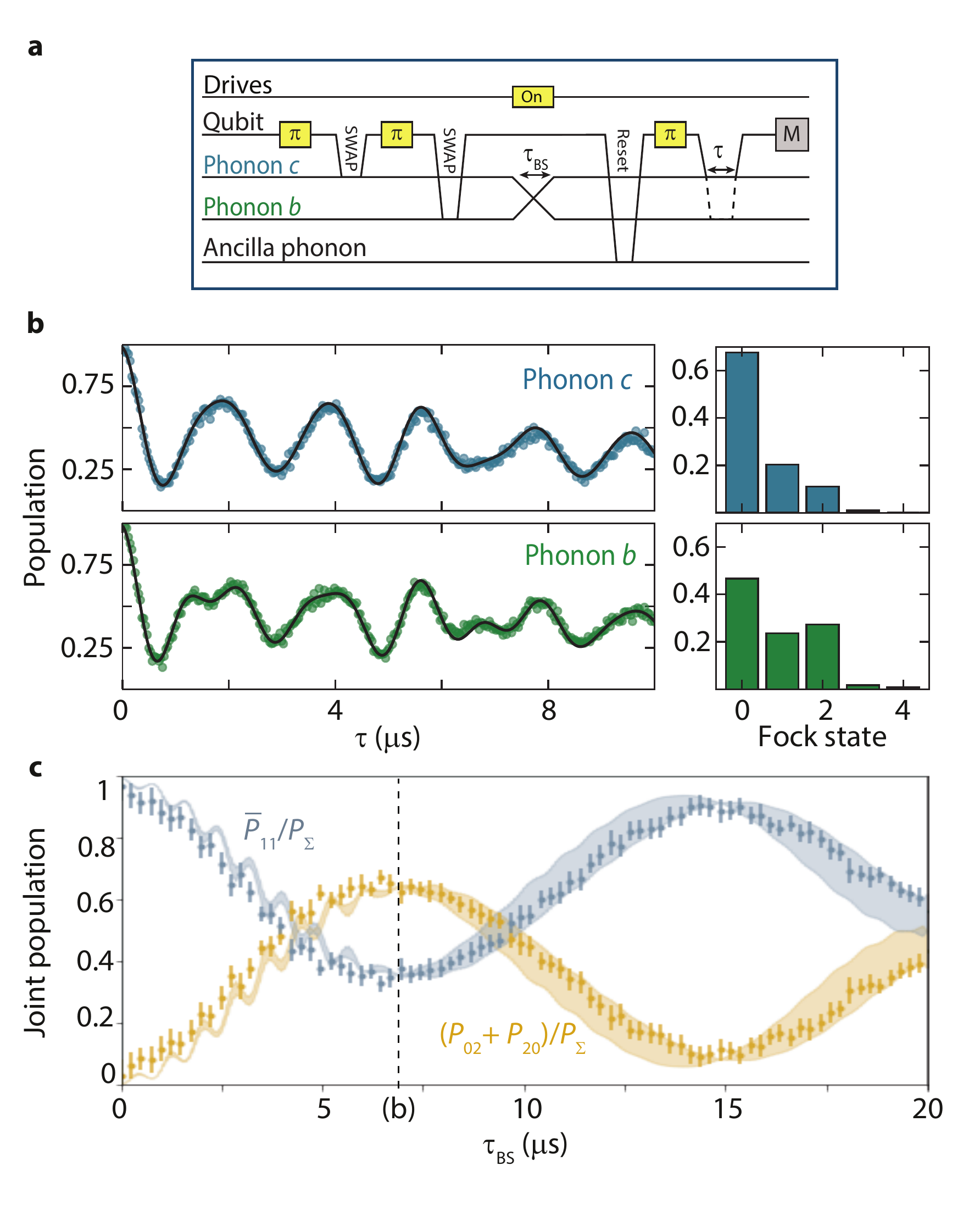}}
    \caption{\textsf{\textbf{Observation of the Hong-Ou-Mandel effect between two phonon modes.} \textbf{a} Pulse sequence used in the experiment. The regime addressed here is the same as for Fig.~\ref{fig:Fig2}, i.e. a two-mode coupling between phonons $c$ and $b$. \textbf{b} Rabi oscillations between phonon modes $c$ ($b$) and the qubit in the top (bottom) plot. Circles are data and black lines are fits.  The extracted Fock state populations for each of the modes is shown in the histograms on the right side. \textbf{c} Normalized joint phonon population for different interaction times $\tau_\textrm{BS}$. Dots are data, shaded areas are the result of simulations that account for 3\% deviation in $g_m$. The error bars on the data points include higher Fock state populations and fit uncertainties, and the dashed vertical line indicates the data shown in \textbf{b}.}}
    \label{fig:Fig4}
\end{figure}

So far, we have studied the two and three mode coupling regimes for the particular case where a single phononic quantum is shared between all participating modes. 
We now investigate the interplay of two quanta during a beam-splitter operation. We first create a $\ket{cb}=\ket{11}$ state in modes $b$ and $c$ by repeatedly exciting the qubit and swapping its excitation into each mode \cite{Chu18}.
We then turn on the two-mode beam-splitter interaction and measure the resulting phonon Fock state distributions of either mode by monitoring the qubit population during a resonant qubit-phonon JC interaction, as shown in previous works \cite{Chu18} (see  Fig.~\ref{fig:Fig4}\textbf{a}). 
As an example, the results for a beam-splitter time of $\tau_\mr{BS}=6.7\,\mu s$ are shown in Fig.~\ref{fig:Fig4}\textbf{b}. 
Here, to optimize the coupling strength and reduce the residual JC-interaction with the qubit, we use a slightly larger qubit-phonon detuning $\tilde{\D}_b=2\pi\cdot 1.2\,$MHz and modulation depth $\Lambda'/\D_{21}=0.85$, resulting in $g_{bc}=2\pi\cdot 18.5\,$kHz. \\

The Hong-Ou-Mandel effect predicts that the outcome of this experiment should depend on whether or not the two phonons are distinguishable.
If they are, no interference between them will occur and the excitations will be shared equally between the two phonon modes. On the other hand, if they are indistinguishable, both excitations will bunch in one of the two phonon modes after the beam-splitter. 
To confirm this experimentally, we compare the probability of the bunched ($P_{20}+P_{02}$) with that of the anti-bunched outcome ($P_{11}$). 
We extract the bunched outcome probability from the individual Fock distributions by assigning $P_{02}+P_{20}$ to $P_2^c+P_2^b$, where $P_2^{c(b)}$ is the probability of finding two quanta in mode $c$~($b$). 
Doing so relies on the assumption that our system contains a maximum of two excitations at the start of the beam-splitter interaction and that no additional quanta are added during the sequence. This assumption is justified because the residual thermal population of the phonon modes is less than $1.6$\% \cite{schrinski_2022}.
Under the same assumption, we can put an upper bound on the anti-bunched probability, namely $\bar{P}_{11} = \min{(P_1^b, P_1^c)} \geq P_{11}$. Nevertheless, we still take into account the possibility for leakage into higher Fock states by fitting the qubit-phonon Rabi oscillations for the first five energy levels. The population contribution of these higher levels is on average 0.01 and is then included in the error bars of Fig.~\ref{fig:Fig4}\textbf{c}.\\

In Fig.~\ref{fig:Fig4}\textbf{c}, we show both $\bar{P}_{11}$ and $P_{20}+P_{02}$ for various beam-splitter interaction times $\tau_\mr{BS}$, normalized by the entire two-excitation subspace $P_\Sigma = P_\textrm{20} + P_\textrm{02} + \bar{P}_{11}$.  As expected, the two-excitation manifold of the phonon state in the beginning of the interaction is dominated by $\ket{11}$. After $\tau_\mr{BS}=6.7\,\mu$s, which corresponds to a 50:50 beam-splitter (vertical dashed line in Fig.~\ref{fig:Fig4}\textbf{c}), the joint state is more likely to be bunched with $(P_{20}+P_{02})/P_\Sigma = 0.622\pm 0.028$. \\

While we cannot straightforwardly access the joint Fock distributions of the two phonon modes in our experiment, we can do so in a master equation simulation of our system using independently measured system parameters. The results are plotted as continuous lines in Fig.~\ref{fig:Fig4}\textbf{c}, showing good agreement between data and theory. The fast oscillations which can be seen for lower interaction times in both theory and experiment arise due to an off-resonant JC-interaction with the qubit. 
This result demonstrates how two a-priori distinguishable phononic quanta in modes at different frequencies are made indistinguishable by a frequency-converting coupling which compensates for the energy difference between the two modes, thereby confirming that the lattice vibrations constituting our phonons display behavior that cannot be described classically. \\ 

In conclusion, we have engineered a direct beam-splitter coupling between two and three distinct mechanical modes of an HBAR. We have used the two-mode interaction to create a phononic $\sqrt{i\mr{SWAP}}$ gate, allowing us to generate entanglement between the modes and observe the Hong-Ou-Mandel effect between two phonons. In addition to our experimental data, we have also presented a theoretical model that is in good agreement with our findings and captures how the single-photon and two-photon conversion processes depend on the drive strength. 
Our results provide a fundamental building block for realizing a QRAM in cQAD by providing one of two crucial operations necessary \cite{Hann2019}, the other one being a conditional phase operation \cite{gao2019}. 
Furthermore, both our current system and many of the concepts discussed here can potentially be applied for the generation of a two-mode squeezing interaction between phonon modes \cite{zhang2019engineering}. Together with single-mode squeezing, these two interactions would enable Gaussian quantum information processing \cite{Weedbrook12} and quantum simulations with bosonic modes \cite{Huh15}.

\subsection*{Acknowledgements}
\noindent The authors thank Ewold Verhagen, Yaxing Zhang, and Marius Bild for useful discussions. Fabrication of the device was performed at the FIRST cleanroom of ETH Z\"urich and the BRNC cleanroom of IBM Z\"urich. We acknowledge support from the Swiss National Science Foundation under grant $200021\_204073$. MF was supported by The Branco Weiss Fellowship -- Society in Science, administered by the ETH Z\"urich. 

\vspace{15pt}

\subsection*{Author contributions}
\noindent U.v.L. designed and fabricated the device. U.v.L., I.C.R., and Y.Y. performed the experiments and analysed the data. U.v.L. developed the theoretical model and performed the QuTiP simulations of the experiments. M.F. provided theory support. Y.C. supervised the work. U.v.L., I.C.R. and Y.C. wrote the manuscript with input from all authors.

\subsection*{Competing interest}
The authors declare no competing interests.

\bibliography{bs.bib}
\clearpage
% \widetext

% %%%%%%%%%%%%%%%%%%%%%%%%%%%%%%%%%%%%%%
% \foreach \x in {1,...,14}
% {%
% \includepdf[pages={\x,{}}]{Supp}
% }

\renewcommand\D{\Delta}
\newcommand\om{\omega}
\newcommand\Om{\Omega}

\renewcommand{\thetable}{S\arabic{table}}  
\renewcommand{\thepage}{S\arabic{page}}  
\renewcommand{\thefigure}{S\arabic{figure}}
\renewcommand{\theequation}{S\arabic{equation}}
\setcounter{page}{1}
\setcounter{figure}{0}
\setcounter{table}{0}
\setcounter{section}{0}
\setcounter{equation}{0}

\widetext

{\centering\textbf{\Large Supplementary Material for} \\} 
{\centering\textbf{\Large Engineering phonon-phonon interactions in multimode circuit quantum acousto-dynamics}\\} 
\normalsize
\vspace{.3cm}
{\centering Uwe von L\"upke, Ines C.~Rodrigues, Yu Yang, Matteo Fadel, and Yiwen Chu\\}
% \vspace{.2cm}

\suppressfloats

\section{Device parameters and measurement setup}
\begin{figure}[tbp]
\centerline{\includegraphics[trim = {0cm, 6cm, 0.0cm, 0.0cm}, clip=True,scale=0.7]{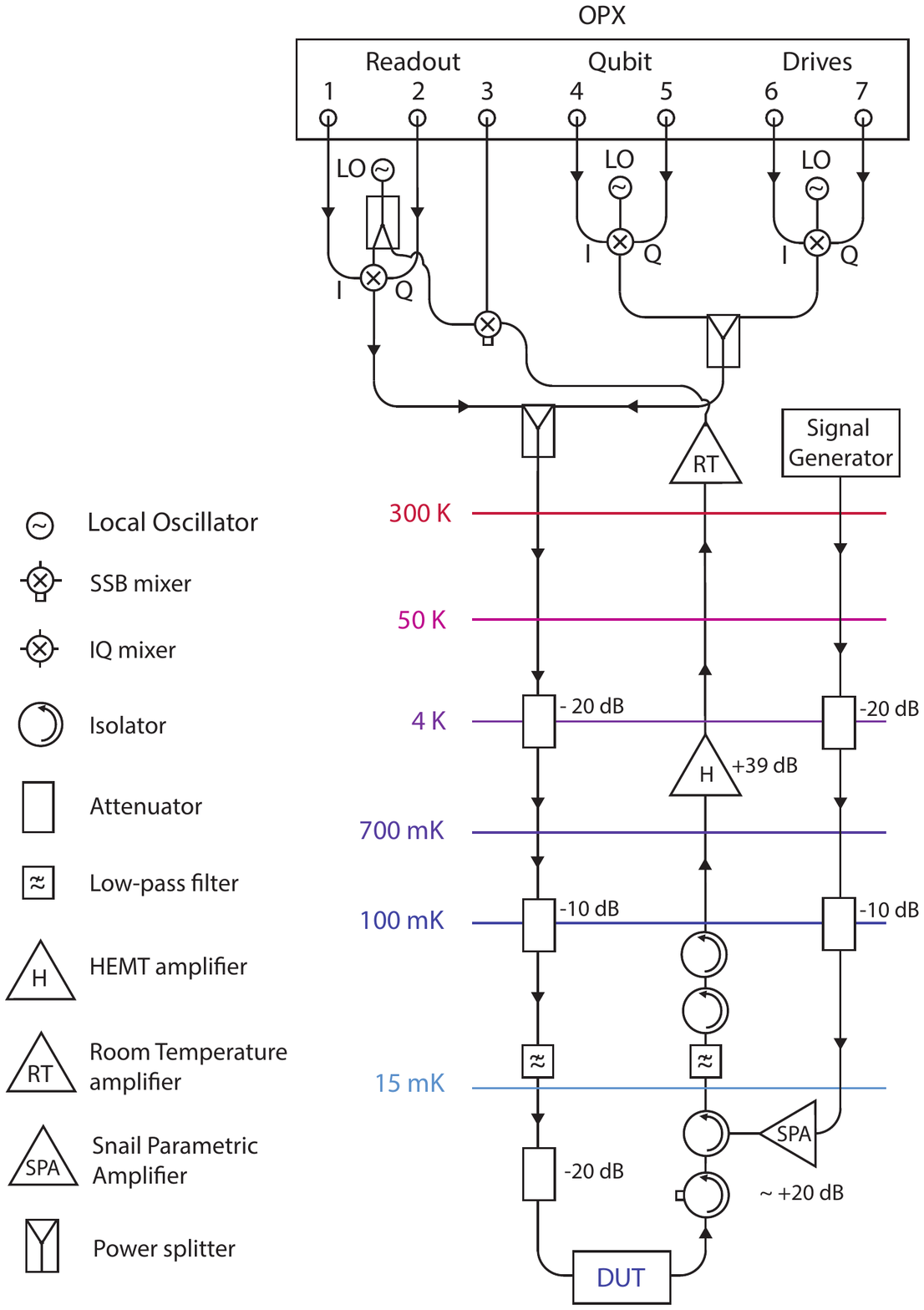}}
\caption{\textsf{\textbf{Wiring diagram.} In the upper part of the figure, we show the room temperature microwave signal setup in which signals are generated by the OPX and upconverted to GHz frequencies by means of IQ mixers. In the lower part we sketch the cabling inside the fridge, including our Device Under Study (DUT) and the parametric amplifier, both mounted in the mK stage of our dilution fridge.}}
\label{fig:wiringDiagram}
\end{figure}

The hybrid device, which we also used for previous studies   \cite{vonLupke22}  \cite{bild2022schr}, consists of a transmon qubit and an HBAR resonator, fabricated on separate chips and flip-chip bonded together. We are using up to four high overtone modes of the HBAR resonator: three of them are used in the experiments shown in the main paper and one ancilla mode is used to reset our transmon qubit. In Table \ref{tab:parameters} we list the parameters of our device, which we obtained through independent measurements. 
The sample sits in a 3D aluminum cavity to both shield it from the environment and to serve as a readout mode for the qubit via the dispersive readout commonly used in cQED. \\
Our microwave cavity is thermalized to the mK stage of a dilution refrigerator and placed inside Mu-Metal shields to shield it from the electromagnetic environment. We generate input signals with a Quantum Machines OPX and upconvert them to GHz frequencies by means of IQ mixers and local oscillators. 
The input lines are thermalized to each stage, resulting in an effective temperature of the qubit which can be measured by probing its residual thermal population, of approximately 50\,mK. Furthermore, we use a SNAIL parametric amplifier \cite{frattini20173} to amplify the readout transmitted signals coming from the output port of the aluminum cavity. 
The amplified readout signal is down-converted using a single-sideband (SSB) mixer and demodulated in the OPX. 
 Fig.~\ref{fig:wiringDiagram} shows both the room temperature signal routing and the wiring inside the fridge. \\
 
\begin{table}[h!]
\begin{tabular}{c|c|c}
\hline
\textbf{Variable}          & \textbf{Parameter} & \textbf{Value}    \\ \hline
$\omega_q$ & qubit frequency without AC Stark shift &  $2\pi\cdot5.971323\,$GHz$\pm 5\,$kHz \\ 
$T_1$ &  qubit relaxation time &  $9.5\pm0.1\,\mu$s \\ 
$T_2^*$ &  qubit coherence time (Ramsey) &  $7.2\pm 0.2\,\mu$s \\ 
$T_2^E$ &  qubit coherence time (Echo) &  $10.3\pm 0.3\,\mu$s \\
$\alpha$ &  qubit anharmonicity &  $2\pi\cdot218\pm0.5\,$MHz\\ 
$\omega_a$ &  resonance frequency of phonon mode $a$ & $2\pi\cdot5.9236\,$GHz$\pm 2\,$kHz \\ 
$\omega_b$ & resonance frequency of phonon mode $b$ &  $2\pi\cdot5.9488\,$GHz$\pm 2\,$kHz \\ 
$\omega_c$ & resonance frequency of phonon mode $c$ & $2\pi\cdot5.9615\,$GHz$\pm 1\,$kHz \\ 
$\Gamma_a$ &  decay rate of phonon mode $a$ &  $2\pi\cdot4.7\pm0.1\,$kHz\\
$\Gamma_b$ &  decay rate of phonon mode $b$ &  $2\pi\cdot3.1\pm0.1\,$kHz\\
$\Gamma_c$ & decay rate of phonon mode $c$ & $2\pi\cdot2.2\pm0.1\,$kHz\\
$g_m$ &  qubit-phonon coupling & $2\pi\cdot257\pm3\,$kHz\\
FSR &  HBAR free spectral range &  $2\pi\cdot 12.62955\,$MHz$\pm3\,$kHz \\ 
\hline
\end{tabular}
\caption{\textsf{\textbf{List of device parameters.}} The errors are the fit uncertainties from the measurements determining the parameters.  }\label{tab:parameters}
\end{table}

\section{Hamiltonian of a bichromatically driven qubit}
\label{section:driven_qubit}
\noindent In this section, we present an analytical description of the qubit behavior when driven with the two parametric drives. In particular, we show the emergence of the qubit sidebands in a time-dependent rotating frame and the expression describing the qubit spectroscopy measurements shown in Fig.~1 of the main text. \\
We start from the Hamiltonian of a transmon qubit, driven by two parametric drives at frequencies $\omega_{1,2}$ and strength $\Omega_{1,2}$, and a weak probe tone 
\begin{equation}
H_\mr{qd} = \omega_q {{q}^\dag} {q}- \frac{\alpha}{2} {{q}^\dag}^{2}{q}^{2} + \left( \Omega_{1} e^{- i \omega_{1} t} {{q}^\dag} + \Omega_{2} e^{- i \omega_{2} t} {{q}^\dag} + \mathrm{h.c.}\right) + \left( \Omega_{p} q^\dag e^{-i\omega t} +\mr{h.c.}\right).
\end{equation}
Here $q,q^\dagger$ are the qubit annihilation and creation operators, $\alpha$ is the qubit anharmonicity and $\Omega_p, \omega$ are the strength and frequency of the probe tone, respectively. In the rotating frame of the qubit g-e transition ($\omega_q$), the Hamiltonian becomes 
\begin{equation}
H_\mr{qd}' = - \frac{\alpha}{2} {{q}^\dag}^{2}{q}^{2} + \left( \Omega_{1} e^{- i \Delta_{1} t} {{q}^\dag} + \Omega_{2} e^{- i \Delta_{2} t} {{q}^\dag} + \mathrm{h.c.} \right)+\left( \Omega_p q^\dag e^{-i\delta_p t} +\mr{h.c.}\right),
\end{equation}
where $\Delta_{1,2}=\omega_{1,2} - \omega_q $ is the detuning between the drive frequencies and the qubit g-e transition and $\delta_p = \omega - \omega_q$ is the detuning between the probe frequency and the qubit g-e transition. 
In the following we assume $\delta_p \ll \Delta_{1,2}$.
Now we apply a unitary transformation 
\begin{equation}
U_{\mathrm{d}} = \exp\left[\xi_1^{*} e^{i \Delta_{1} t} {q} + \xi_2^{*} e^{i \Delta_{2} t} {q} - \mathrm{h.c.}  \right]
\label{eq.Ud}
\end{equation}
to move to the interaction picture, also known as the displaced frame, of these drives, where $\xi_{1,2}=\Omega_{1,2}/\Delta_{1,2}$. 
This transforms the qubit operator as 
\begin{equation}
q'=U_\mr{d}qU_\mr{d}^\dag = \xi_{1} e^{- i \Delta_{1} t} + \xi_{2} e^{- i \Delta_{2} t} + {q}.
\end{equation}
From now on we will assume that $\xi_1$ is real-valued and write $\xi_2$ as $\xi_2 e^{-i\phi}$ with $\xi_2\in\mathbb{R}$. Under this assumption, $\phi$ describes the initial phase difference between the two drives. This notation serves to avoid complex drive constants and highlights the effect of an initial phase difference. 
While interesting physics can be studied when varying $\phi$ \cite{Pino22}, all the experiments presented in the main text are done using $\phi=0$. 
Furthermore, we define $\Delta_{21}\equiv\Delta_2-\Delta_1$, $\Sigma_{21}=\Delta_1 + \Delta_2$, and choose $\Delta_{21}>0$. 
The Hamiltonian $H_\mr{qd}'$ then transforms as 
\begin{align}
H_\mr{qd}'' &= U_{d}H_\mr{qd}' U_{d}^\dag + i\dot{U}_\mr{d}{U}_\mr{d}^\dag \label{eq:drive-interaction-picture} \\
 &=- \frac{\alpha}{2} {{q}^\dag}^{2} {q}^{2}\nonumber\\ 
 &+ \alpha \left(- \xi_{1} e^{- i \Delta_{1} t} - \xi_{2} e^{- i \phi} e^{- i \Delta_{2} t}\right) {{q}^\dag}^{2} {q}+\mr{h.c.}\nonumber\\ 
 &+ \alpha \left(- 2 \xi_{1}^{2} - 2 \xi_{1} \xi_{2} e^{i \phi} e^{i \Delta_{21} t} - 2 \xi_{1} \xi_{2} e^{- i \phi} e^{- i \Delta_{21} t} - 2 \xi_{2}^{2}\right) {{q}^\dag} {q}\nonumber\\ 
 &+ \alpha \left(- \frac{\xi_{1}^{2}}{2} e^{- 2 i \Delta_{1} t} - \xi_{1} \xi_{2} e^{- i \phi} e^{- i \Sigma_{21} t} - \frac{\xi_{2}^{2}}{2} e^{- 2 i \phi} e^{- 2 i \Delta_{2} t}\right) {{q}^\dag}^{2}+\mr{h.c.}\nonumber\\ 
 &+ \alpha \big(- \xi_{1}^{3} e^{- i \Delta_{1} t} - \xi_{1}^{2} \xi_{2} e^{i \phi} e^{- 2 i \Delta_{1} t} e^{i \Delta_{2} t} - 2 \xi_{1}^{2} \xi_{2} e^{- i \phi} e^{- i \Delta_{2} t}\nonumber \\
 &\qquad - 2 \xi_{1} \xi_{2}^{2} e^{- i \Delta_{1} t} - \xi_{1} \xi_{2}^{2} e^{- 2 i \phi} e^{i \Delta_{1} t} e^{- 2 i \Delta_{2} t} - \xi_{2}^{3} e^{- i \phi} e^{- i \Delta_{2} t}\big) {{q}^\dag} +\mr{h.c.} \nonumber\\
 &+\left( \Omega_p q^\dag e^{-i\delta_p t} +\mr{h.c.}\right). \label{eq:Hqd_pp}
\end{align}
Now we will drop all terms oscillating faster than $\Delta_{21}$ in a Rotating Wave Approximation (RWA). Note that $\delta_p \simeq \Delta_{21}$, such that we keep the term in the last line of Eq. (\ref{eq:Hqd_pp}). 
Since $\Delta_{21}<\Delta_1<\Delta_2<\Sigma_{21}$ in our setup, we can drop the terms proportional to the non-diagonal operators ${{q}^\dag}^{2} {q}$, ${{q}^\dag}^{2}$, ${{q}^\dag}$ and h.c., except for the probe. 
Nevertheless, we emphasize that the relation between the different detunings can be modified for a different choice of resonance condition, which could make some of the terms we dropped in this RWA more relevant in other parameter regimes than the ones chosen in this work. \\

This RWA leads to a Hamiltonian diagonal in the qubit operator, namely
\begin{align}
H_{\mr{RWA}} &=- \frac{\alpha }{2} {{q}^\dag}^{2}{q}^{2}+  \left(- 2 \alpha\xi_{1}^{2} - 2 \alpha\xi_{2}^{2} - 2\alpha \xi_{1} \xi_{2} e^{i \phi} e^{i \Delta_{21} t} - 2\alpha \xi_{1} \xi_{2} e^{- i \phi} e^{- i \Delta_{21} t} \right) {{q}^\dag} {q} \nonumber \\ 
&\quad +\left( \Omega_p q^\dag e^{-i\delta_p t} +\mr{h.c.}\right)\\
&= - \frac{\alpha }{2} {{q}^\dag}^{2}{q}^{2}+  \left(\underbrace{- 2 \alpha\xi_{1}^{2} - 2 \alpha\xi_{2}^{2}}_{\delta_q^{ss}} \underbrace{- 4\alpha \xi_{1} \xi_{2}}_{\Lambda} \cos{(\Delta_{21}t+\phi)}\right) {{q}^\dag} {q}+\left( \Omega_p q^\dag e^{-i\delta_p t} +\mr{h.c.}\right). 
\label{eq.Hrwa1}
\end{align}
In Eq.~(\ref{eq.Hrwa1}) the expression in large brackets represents the Stark shift of the qubit due to the presence of the parametric drives, also shown as Eq. (2) in the main text. 
This term has a time-independent part $\left(\delta_q^{ss}\right)$ and a time-dependent part (term with $\Lambda$ as prefactor), the former acting as a frequency shift of the qubit and the latter as a modulation of the qubit frequency. 
The modulation has a modulation frequency $\D_{21}$ and modulation depth $\Lambda / \D_{21}$.
We move our qubit rotating frame to that of the Stark shifted qubit, $q\rightarrow q e^{-i\delta_q^{ss}t}$, which removes the constant frequency shift $\delta_q^{ss}$ and modifies the detuning of the probe tone to $\tilde{\delta}_p = \delta_p - \delta_q^{ss}$. 
Note that $\delta_q^{ss}$ acquires a correction in the parameter regime we work in, which is shown in Section \ref{section:drive-calibration}. 
Entering this rotating frame also has an effect on coupling terms, which we will introduce later when we treat the phonon modes. 
In this frame the qubit Hamiltonian is
\begin{equation}
H_\mr{RWA}' =  - \frac{\alpha }{2} {{q}^\dag}^{2}{q}^{2} - 4\alpha \xi_{1} \xi_{2} \cos{(\Delta_{21}t+\phi)} {{q}^\dag} {q}+\left( \Omega_p q^\dag e^{-i\tilde{\delta}_p t} +\mr{h.c.}\right)~. \label{eq.Hrwa2}
\end{equation}
Next we move into the time-dependent rotating frame of the frequency modulation by applying the transformation
\begin{equation}
U_{\mr{JA}} = \exp\left[i\frac{\Lambda}{\Delta_{21}} \sin{(\D_{21}t+\phi)} {q}^\dag{q} \right]. \label{eq.JA-trafo}
\end{equation}
This removes the time-dependent qubit frequency term in Eq.~(\ref{eq.Hrwa2}) since 
\begin{equation}
i\dot{U}_{\mr{JA}}{U}_{\mr{JA}}^\dag = -\Lambda\cos{(\Delta_{21}t+\phi)} q^\dag q
\end{equation} 
and transforms the qubit operator according to the Jacobi-Anger expansion as
\begin{equation}
q_{\mr{JA}} = {q} \exp\left[- i\frac{\Lambda}{\Delta_{21}} \sin{ (\Delta_{21} t + \phi )}\right] = {q} \sum_{n=-\infty}^{\infty}J_n\left(\frac{\Lambda}{\D_{21}}\right) e^{-in(\D_{21}t+\phi)}, 
\end{equation}
where $J_n$ is the Bessel function of the first kind and $n$ refers to the sideband number. 

Since this frame transformation leads to a time-dependent phase in the qubit operator, it does not affect the diagonal terms in $H_{\mr{RWA}}$, but it does affect the probe term as we can see in the transformed Hamiltonian
\begin{equation}
H_\mr{JA} = - \frac{\alpha }{2} {{q}^\dag}^{2}{q}^{2}+ \left( \Omega_p {q}^\dag \sum_{n=-\infty}^{\infty}J_n\left(\frac{\Lambda}{\D_{21}}\right) e^{in(\D_{21}t+\phi)} e^{-i\tilde{\delta}_p t} +\mr{h.c.}\right).
\label{eq:driven_qubitH}
\end{equation}
By noting that $\Omega_p$ is small compared to $\D_{21}$ we can perform an additional RWA to keep only time-independent terms. 
Thus, we keep only the terms which satisfy the condition $n\D_{21}=\tilde{\delta}_p$, leading to the interpretation that the probe only affects the qubit when it is detuned from the Stark shifted g-e transition by integer multiples of $\Delta_{21}$. An experimental confirmation of this can be seen through the appearance of multiple qubit "sidebands" during qubit spectroscopy (see Fig.~1\textbf{b} of the main paper). 

\subsection*{Qubit response}
To model the spectroscopic response of our transmon qubit when driven by two tones and probed by a third weak tone we make use of the Bloch equations in steady state, and write the excited state population \cite{schuster2005ac} as

\begin{equation}
P_e = \frac{1}{2} \sum_{n=-\infty}^{\infty} \frac{\Omega_n^2T_1T_2^*}{1+(T_2^*\Delta)^2 + \Omega_n^2T_1T_2^*}.
\label{eq:qubitPe}
\end{equation}

Here $T_1$ is the energy relaxation time, $T_2^*$ is the Ramsey decoherence time and $\Delta = \omega - \omega_q - n\Delta_{21}$ is the detuning between the probe tone and the generated qubit sidebands. Furthermore, the relative probe strength can be expressed as

\begin{equation}
\Omega_n = J_n\left(\frac{\Lambda}{\D_{21}}\right) \Omega_p.
\end{equation}
We use Eq.~(\ref{eq:qubitPe}) to model the data shown in Fig.~1 of the main paper.

\section{Beam-splitter interaction Hamiltonian}\label{sec:beamsplitterHamiltonian}
So far we have seen what happens to the Hamiltonian of a bichromatically driven qubit when probed by a weak probe tone. 
In this section, we will study the effect of the driven qubit on a finite number of harmonic oscillators, for example two phonon modes $a$ and $b$.\\
The initial Hamiltonian in the lab frame and without the probe tone is given by
\begin{equation}
H_\mr{JC} = \omega_q {{q}^\dag} {q}- \frac{\alpha}{2} {{q}^\dag}^{2}{q}^{2} + \left( \Omega_{1} e^{- i \omega_{1} t} {{q}^\dag} + \Omega_{2} e^{- i \omega_{2} t -i\phi} {{q}^\dag} + \mathrm{h.c.}\right)  + \omega_a {{a}^\dag} {a} + \omega_b {{b}^\dag} {b}+ \left(g_a a^\dag q + g_b b^\dag q +\mr{h.c.}\right) \label{eq:HJC}
\end{equation}
Now we apply the same transformations as in Section \ref{section:driven_qubit} and, in addition, also move the phonons to a frame rotating with their resonance frequencies $\om_{a,b}$. 
Right before we enter the time-dependent rotating frame with the transformation $U_\mr{JA}$ in Eq.~(\ref{eq.JA-trafo}), the Hamiltonian is
\begin{align}
H_\mr{RWA}^\mr{m} =& - \frac{\alpha }{2} {{q}^\dag}^{2}{q}^{2} \underbrace{- 4\alpha \xi_{1} \xi_{2}}_{\Lambda}\cos{(\Delta_{21}t+\phi)}{{q}^\dag} {q} \nonumber \\
& + \left(g_a a^\dag q e^{i\tilde{\D}_{a}t} + g_b b^\dag q  e^{i\tilde{\D}_{b}t} +\mr{h.c.}\right) \nonumber \\
& +\left( \big(g_a a^\dag e^{i\D_a t} + g_b b^\dag e^{i\D_b t}\big) \big( \xi_1 e^{-i\D_1 t} + \xi_2 e^{-i\D_2 t - i\phi}\big)+\mr{h.c.}\right), 
\label{eq.Hmrwa}
\end{align}
with $\tilde{\D}_{a,b} = \omega_{a,b} - (\omega_q + \delta_q^{ss}) $ and the modulation depth $\Lambda$ as in Section \ref{section:driven_qubit}. \\

The second line of Eq.~(\ref{eq.Hmrwa}) describes the off-resonant coupling between qubit and phonon modes and the third line represents an effective drive on the phonon modes mediated by the qubit. 
Note that the strength of the effective drive depends on the detuning $\D_{a,b}$ between the bare qubit and the phonon frequencies as it is the result of the transformation $U_\mr{d}$ in Eq.~(\ref{eq.Ud}). 
Its resonance condition depends on the phonon-drive detuning $\D_{a,b}-\D_{1,2}$, which is large for the parametric drives considered here, such that we drop the effective drive terms. 

Now applying $U_\mr{JA}$ and also using the shorthand $J_n(\frac{\Lambda}{\D_{21}})\equiv J_n$ for better readability, we get 
\begin{align}
H_\mr{JA}^\mr{m} = & \underbrace{- \frac{\alpha }{2} {{q}^\dag}^{2}{q}^{2} }_{H_\mr{Kerr}} + \underbrace{\left(g_a a^\dag q \sum_{n}J_n e^{i( \tilde{\Delta}_{a}  - n\Delta_{21} ) t } e^{-in\phi} + g_b b^\dag q \sum_{k}J_k e^{i( \tilde{\Delta}_{b}  - k\Delta_{21} ) t } e^{-ik\phi} +\mr{h.c.}\right)}_{H_\mr{sideband~coupling}= V}, \label{eq.HmJA}
\end{align} 
where both sums run from $-\infty$ to $\infty$. Note that for better readability, we use the indices $n$ and $k$ to represent the sideband number for the $a^\dagger q$ term and the $b^\dagger q$ term. 

From the second term of Eq.~(\ref{eq.HmJA}), we see that the qubit sidebands (which are generated by the frequency modulation arising from the two drives) couple individually to the phonon modes. 
This can be used to activate and control a coupling between the qubit and a detuned phonon mode $m$ with a coupling $g_m J_n$ by varying $\Lambda$ and $\D_{21}$ \cite{Strand13, Naik2017}.

Moreover, the second line of Eq.~(\ref{eq.HmJA}) can be understood as an off-resonant coupling between the qubit and the two phonon modes ($a$ and $b$) through each sideband ($J_n$ and $J_k$). 
The coupling via the $n(k)^{th}$ sideband comes with a phase $\tilde{\Delta}_{a(b)}  - n(k)\Delta_{21}$, such that most of the infinite sum terms are fast oscillating. 
The sideband closest to the respective phonon mode is detuned only by about 1\,MHz in our experiment, such that it leaves a small residual JC-interaction between qubit and phonon.\\ 

We now move into an interaction picture of the coupling between the qubit and the phonon modes through the different sidebands via the transformation 
\begin{equation}
U_{c} = \exp\bigg[\sum_{n} \frac{g_{a}J_{n}}{\tilde{\Delta}_{a}  - n\Delta_{21}} {{a}^\dag} {q} e^{i( \tilde{\Delta}_{a}  - n\Delta_{21} ) t } e^{-in\phi} + \sum_{k} \frac{g_{b} J_{k}}{\tilde{\Delta}_{b}  - k\Delta_{21}} {{b}^\dag} {q} e^{i( \tilde{\Delta}_{b}  - k\Delta_{21} ) t } e^{-ik\phi}   -\mr{h.c.}\bigg] \equiv e^{S}.
\end{equation}

Note that this is analogous to the usual Schrieffer-Wolff transformation used in the dispersive regime of a qubit and resonator coupled through the Jaynes-Cummings interaction. The transformed Hamiltonian is 
\begin{align}
H_c &= U_c H_\mr{JA}^\mr{m} U_c^\dag  + i \dot{U}_\mr{c} U_\mr{c}^\dag \\
&= U_c H_\mr{Kerr} U_c^\dag + U_c V U_c^\dag + i \frac{\partial }{\partial t}\left( e^S\right) e^{-S} \label{eq:Hc}
\end{align}
$U_c H_\mr{Kerr} U_c^\dag$ yields terms comprised of a total of four phonon ($a,b$) and qubit ($q$) operators with appropriate prefactors. 
The photon number dependence of the qubit transition frequency described by $H_\mr{Kerr}$ leads to a modification of the effective drive strengths and of the modulation depth, which we describe in the following section \cite{Koch2007}. 
This modification also applies to the effective beam-splitter coupling rates, but we neglect this effect in our analysis as it is negligible if $\D_{a,b}\ll\alpha$. Note that this is not the case for some previous works \cite{gao2019}, where the correction would have to be taken into account to correctly predict the coupling rates and frequency shifts. 
Furthermore, $H_\mr{Kerr}$ creates combinations of phonon and qubit operators with drive-related terms arising from the transformation in Eq.~(\ref{eq:drive-interaction-picture}). 
These forth order terms become relevant when their respective resonance condition is met and their amplitudes are large enough. 
Their effects are both interesting for further studies and troublesome when overshadowing the physics we want to highlight in this work. Even though for the experiments presented here we avoided hitting these resonances, they might pose a challenge when scaling up to a larger number of simultaneously driven interactions. \\
Our main focus is the interaction term $V$ and how the transformation $U_c$ reveals the beam-splitter interaction between the phonon modes. 
To that end, we use the Baker-Campbell-Hausdorff formula to explicitly execute the above transformation. 
Noting that $i\partial S/\partial t =-V$ and expanding the derivative of $e^S$ as a Taylor series, we can write the effect of $V$ under the coupling interaction picture $U_c$ as 
\begin{align}
& U_c V U_c^\dag + i \frac{\partial }{\partial t}\left( e^S\right) e^{-S} \label{eq:UVUd}\\
=& \sum_{j=0}^\infty \frac{1}{j!}\left(\mr{ad}_S \right)^j V +i \sum_{j=0}^\infty \frac{1}{(j+1)!}\left(\mr{ad}_S \right)^j \frac{\partial S}{\partial t} \label{eq:adS}\\
= & V + [S,V] +i\frac{\partial S}{\partial t} +i\frac{1}{2}\left[S,\frac{\partial S}{\partial t}\right] + \mathcal{O}\left(g_{a,b}^3/\D_{a,b}^3\right)  \\
= & V + [S,V] -V -\frac{1}{2}[S,V] + \mathcal{O}\left(g_{a,b}^3/\D_{a,b}^3\right) \\
=& \frac{1}{2}[S,V] \label{eq:CommutatorSV}\\
=& \underbrace{\sum_{n,k} \frac{g_{a}^2 J_n J_{k} }{\tilde{\Delta}_{a}  - n\Delta_{21}}   \cos{\left[ (k - n)\Delta_{21}  t + (k-n)\phi\right]} }_{\delta_a} a^\dag a\label{eq.phAss}\\
+& \underbrace{\sum_{n,k} \frac{g_{b}^2 J_n J_{k} }{\tilde{\Delta}_{b}  - k\Delta_{21}}   \cos{\left[ (k - n)\Delta_{21}  t + (k-n)\phi\right]} }_{\delta_b} b^\dag b\label{eq.phBss}\\
-& \underbrace{\sum_{n,k} \left\{\frac{g_{a}^2 J_n J_{k} }{\tilde{\Delta}_{a}  - n\Delta_{21}} + \frac{g_{b}^2 J_n J_{k} }{\tilde{\Delta}_{b}  - k\Delta_{21}} \right\}   \cos{\left[ (k - n)\Delta_{21}  t + (k-n)\phi\right]} }_{\delta_q} q^\dag q\label{eq.qb_shift} \\
+& \underbrace{\sum_{n,k} \frac{1}{2}\left\{\frac{g_a g_{b} J_n J_{k} }{\tilde{\Delta}_{b}  - k\Delta_{21}}+\frac{g_a g_{b} J_n J_{k} }{\tilde{\Delta}_{a}  - n\Delta_{21}}\right\}e^{i(k - n)\Delta_{21} t}e^{i(\Delta_{a}-\D_b) t}e^{i(k-n)\phi} }_{g_\mr{BS} } a^\dag b  + \mr{h.c.}\label{eq.gBS}
\end{align}
where $\mr{ad_S}~\bullet \equiv [S,~ \bullet ]$ in Eq. (\ref{eq:adS}) denotes the adjunct action. A thorough derivation of Eqs. (\ref{eq:UVUd}) to (\ref{eq:CommutatorSV}) can be found in the appendix of Ref. \cite{xiao2022perturbative}. 
Eq.~(\ref{eq.phAss}) and Eq.~(\ref{eq.phBss}) contain phonon frequency shifts $\delta_{a,b}$ due to the presence of the qubit and its sidebands, with Eq. (\ref{eq.qb_shift}) containing an equal, but opposite shift of the qubit frequency. 
As discussed in Ref.~\cite{Gely2021}, these phonon frequency shifts are predominantly due to the normal mode splitting of qubit and phonon modes when approaching an avoided crossing. 
The qubit frequency shifts by the same amount in the opposite direction, evident from Eq.~(\ref{eq.qb_shift}). 
Finally, Eq.~(\ref{eq.gBS}) unveils the phonon-phonon beam-splitter coupling. \\
We can formally eliminate the phonon frequency shifts by entering a rotating frame for the phonon modes which cancels the shifts and adds a corresponding phase to the operators $a$ and $b$. 
This modifies the resonance condition of Eq.~(\ref{eq.gBS}) to $(n-k)\D_{21}=\D_a - \D_b + \delta_a - \delta_b$, similar to how the qubit stark shift modified the qubit-phonon detunings from $\Delta_{a,b}$ to $\tilde{\D}_{a,b}$
In other words, the difference between the two drive frequencies now has to match the difference between the shifted phonon frequencies. 
To realize this modified resonance condition in the experiment we need to adapt $\D_{21}$, which slightly changes the prefactor in Eq. (\ref{eq.gBS}) as we change the denominator. 
However, we can safely assume that $|\tilde{\D}_b - k \D_{21}|\gg |\delta_{a,b}|$, so this change will be small.\\ 

\begin{figure}[tbp]
\centerline{\includegraphics[width=0.8\textwidth]{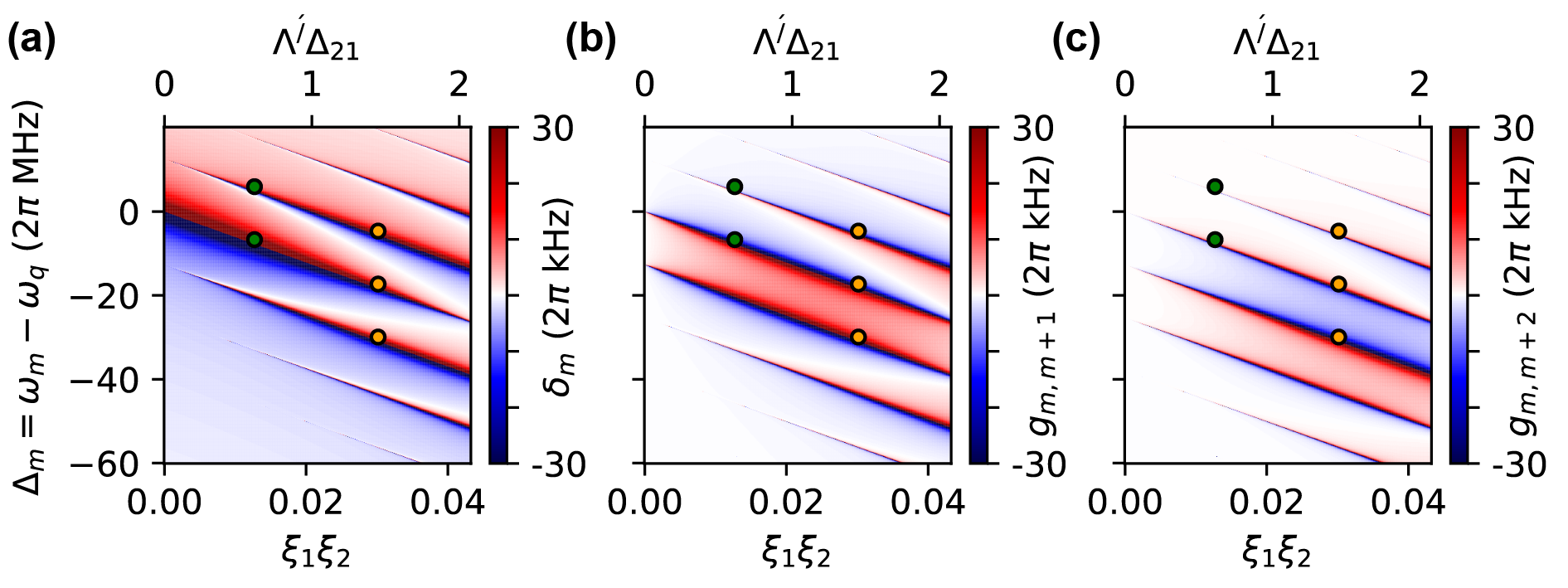}}
\caption{\textsf{\textbf{Analytical result for phonon frequency shift and beam-splitter coupling.} \textbf{a} Frequency shift of a phonon mode $m$ detuned from the qubit by $\Delta_m = \omega_m - \omega_q$ for a drive power $\xi_1\xi_2$. Green (orange) circles indicate the parameters of the 2 (3) phonon modes used in the experiments presented in Fig.~2 (3) of the main text. \textbf{b} Beam-splitter coupling of the same mode to its first higher frequency neighbor $m+1$ at $\Delta_m+\D_{21}$. \textbf{c} Beam-splitter coupling from mode $m$ to its second higher frequency neighbor $m+2$ at $\Delta_m+2\D_{21}$. }}
\label{fig:SItheory}
\end{figure}

Even after eliminating the phonon frequency shifts, we are still left with an unwieldy term for the beam-splitter coupling, which contains two infinite sums and multiple phase factors. 
Thus, to simplify the Hamiltonian further, we make use of the four-wave mixing resonance condition, noting that the phase of the term should remain constant in time. This allows us to write a condition for $k$ in terms of $n$ to eliminate the sum over $k$ since both the phonon-phonon detuning $\D_a-\D_b$ and the drive frequency difference $\D_{21}$ are given by the experiment. 
In the main text, we are using in particular the phonon frequency shifts as well as the coupling terms that arise between neighboring phonon modes when $n=k+1$ and next-to-neighboring phonon modes when $n=k+2$. 
These three terms are simplifications of Eq. (\ref{eq.phAss}) and (\ref{eq.gBS}), using $\phi = 0$ and $(n-k)\D_{21}=\D_a - \D_b$, leading to Eq. (3), (4), and (5) of the main text, which we repeat here for a phonon mode $m$, detuned from the qubit by $\Delta_m$
\begin{equation}
    \delta_m = g_m^2\sum_{n} \frac{ J_n^2 }{\Delta_{m}  - n\Delta_{21}}
    \label{eq:deltam} ,  
\end{equation}
\begin{equation}
    g_\mr{BS}^{m,m+1} = g_{m}g_{m,m+1}  \sum_{n} \frac{ J_n J_{n+1} }{\Delta_{m}  - n\Delta_{21}} 
    \label{eq:gmp1}, 
\end{equation}
\begin{equation}
    g_\mr{BS}^{m,m+2} =  g_{m}g_{m,m+2} \sum_{n} \frac{ J_n J_{n+2} }{\Delta_{m}  - n\Delta_{21}} 
    \label{eq:gmp2}.
\end{equation}
To illustrate how Eq. (\ref{eq:deltam}), (\ref{eq:gmp1}), and (\ref{eq:gmp2}) behave for various drive strengths and phonon-qubit detunings, we plot them in Fig.~\ref{fig:SItheory}. 
The diagonal lines correspond to qubit frequencies, where the denominator $\Delta_m - n\Delta_{21}$ diverges. Note that we omitted the Stark shift correction of $\D_{m}$, so that the effect of the Stark shift can be seen in the downward slope of the features in Fig. \ref{fig:SM_Fig2}. The green (orange) circles indicate the parameters of the phonon modes used in the experiments presented in Fig.~2 (3) of the main text. \\ 

\noindent Lastly, we note that to first order in the BCH-expansion, the qubit and phonon operators transform as 
\begin{align}
U_\mr{c}qU_\mr{c}^\dag &=q + \sum_n \frac{g_{a}J_{n} }{\tilde{\Delta}_{a}  - n\Delta_{21}} {a} e^{-i( \tilde{\Delta}_{a}  - n\Delta_{21} ) t } e^{in\phi}+ \sum_k \frac{g_{b} }{\tilde{\Delta}_{b}  - k\Delta_{21}} {b} e^{-i( \tilde{\Delta}_{b}  - k\Delta_{21} ) t } e^{ik\phi} \\
U_\mr{c}aU_\mr{c}^\dag &= a - \sum_n \frac{g_{a}J_{n} }{\tilde{\Delta}_{a}  - n\Delta_{21}} {q} e^{i( \tilde{\Delta}_{a}  - n\Delta_{21} ) t } e^{-in\phi} \\
U_\mr{c}bU_\mr{c}^\dag &= b - \sum_k \frac{g_{b}J_{k} }{\tilde{\Delta}_{b}  - k\Delta_{21}} {q} e^{i( \tilde{\Delta}_{b}  - k\Delta_{21} ) t } e^{-ik\phi}
\end{align}
These transformations provide a more intuitive understanding of the possible combinations of phonon and qubit operators that can arise when transforming $H_\mr{Kerr}$ and $H_\mr{drive}$. Essentially, the qubit partially hybridizes with all phonon modes through the sideband mediated interaction, allowing us to replace the qubit and phonon operators with the respective hybridized operators, similar to what is done in a standard Schrieffer-Wolff transformation.

\section{Modulation depth correction}
In Section \ref{section:driven_qubit} we saw how going into the interaction picture of the two parametric drives leads to a time-dependent Stark shift on the qubit. 
The resulting time-dependent term acts like a frequency modulation, creating sidebands whose amplitude is proportional to the Bessel function of the first kind. 
The argument of the Bessel functions is the modulation depth $\Lambda/\D_{21}$, where $\Lambda=-4\alpha\xi_1\xi_2$ is the prefactor of the cosine modulation of the qubit frequency. 
However, when taking into account higher energy levels of the qubit, its time-dependent response to the parametric drives becomes more complex, in a similar fashion to how the time-independent stark shift depends on the qubit anharmonicity in Eq.~(\ref{eq:qubitshift}). This is the result of ignoring the Kerr-term in Eq. (\ref{eq:Hc}). 
Here, we use time-independent perturbation theory to derive a correction to the modulation depth due to this effect.
The calculation of the correction is straightforward for a single mode where we can enter a rotating frame at the drive frequency, thereby eliminating the explicit time dependence of the drive term as done for example in \cite{zhang2019engineering}. 
However, this is not obvious for our Hamiltonian which includes two drives. 
To mitigate this, we model our drives as harmonic modes with resonance frequencies $\omega_{1,2}$. These are described by the operators $d_{1,2}^{(\dag)}$ and are coupled to the qubit with coupling strengths $g_{1,2}$. 
The Hamiltonian of the qubit and drives then reads
\begin{equation}
    H=H_0 + \lambda V~,
\end{equation}
where
\begin{align}
    H_0 &= \omega_q q^\dag q - \frac{\alpha}{2}{{q}^\dag}^{2}{q}^{2} + \omega_1 d_1^\dag d_1+\omega_2 d_2^\dag d_2 \\
    V &= g_1d_1^\dag q + g_2d_2^\dag q + \mr{h.c.}
\end{align}
Keeping explicit field terms $d_{1,2}^\dag d_{1,2}$ eliminates the explicit time dependence. 
We now write the joint state of qubit and drives as $\ket{nlm}$, where the indices $n,l,m$ are the photon numbers in qubit ($n$) and the two drive modes ($l,m$). 
By following standard perturbation theory we find the unperturbed energy levels $E^{(0)}_{nlm}$  
\begin{equation}
H_0\ket{nlm}=\left(\omega_q n - \frac{\alpha}{2} n (n-1) + \omega_1 l + \omega_2 m\right)\ket{nlm} \equiv E^{(0)}_{nlm}\ket{nlm}~,
\end{equation}
with the first order correction  
\begin{equation}
E^{(1)}_{nlm} = \bra{nlm}V\ket{nlm}=0~,
\end{equation}
and the second order correction
\begin{equation}
E^{(2)}_{nlm} = \sum_{k\neq n} \frac{|\bra{kst}V\ket{nlm}|^2}{E^{(0)}_{nlm}-E^{(0)}_{kst}}~.\label{eq:correction}
\end{equation}
Note that this second order correction to the energy levels contains many terms which we handle in Wolfram Mathematica and do not reproduce here. 
Nevertheless, with this long but analytically tractable expression for $E^{(2)}_{n}$ in hand, we compute the correction to the transition between neighboring levels $E^{(2)}_{n}-E^{(2)}_{n-1}$.
After evaluating Eq. (\ref{eq:correction}) for our Hamiltonian, we identify the "coupling strengths" as being given by the drive amplitudes scaled by the square root of the photon number in each drive $g_{1,2}\rightarrow \Omega_{1,2}/\sqrt{l,m}$.  This allows us to eliminate the drive occupations $l,m$ from the result.
Next, we identify terms proportional to $\Omega_1 \Omega_2^\star +\mr{h.c.}$ as those  contributing to the frequency modulation and terms proportional to $|\Omega_{1,2}|^2$ as those contributing to the Stark shift. 
Focusing solely on the modulation depth correction we finally arrive at the correction for the lowest qubit transition
\begin{equation}
E^{(2)}_{1}-E^{(2)}_{0} = -2\alpha\Omega_1\Omega_2\left( \frac{1}{\D_1(\D_1+\alpha)}+\frac{1}{\D_2(\D_2+\alpha)}\right)\equiv \Lambda'~, \label{eq:modAmp-correction}
\end{equation}
where $\D_{1,2}=\omega_{1,2}-\omega_q$ as in the previous sections. 
In the figures in the main text, we use this corrected prefactor of the qubit frequency modulation to translate the drive amplitudes $\Omega_{1,2}$ into a corrected modulation depth $\Lambda'/\D_{21}$. Furthermore, note that we take into account the Stark shift correction separately in Section \ref{section:drive-calibration}.

\section{Drive calibration}\label{section:drive-calibration}

We want to calibrate the ratio $\eta_{1,2}$ between the effective drive amplitudes $\Omega_{1,2}$ reaching the qubit and the unitless amplitudes $\Omega_{1,2}^\mr{DAC}$ we set for the digitally created waveforms. 
To that end we perform a spectroscopy measurement where for varying values of $\Omega_{1,2}^\mr{DAC}$, we extract the qubit's Stark-shifted frequency through a Lorentzian fit. We perform this experiment for one drive at a time  while the other one is turned off. 
Since the drives are spaced by $\Delta_{21} = 2\pi \cdot 12.62955$ MHz ($\omega_2>\omega_1$) the power dependent frequency shift of the qubit will be different for each of the tones. The frequency shift is given by an expression analogous to that for the dispersive shift between a transmon qubit and a 3D cavity found in Ref.~\cite{Koch2007}:

\begin{equation}
{\delta_q^{ss}}^{'} = -2|\Omega_{1,2}|^2\frac{\alpha}{\Delta_{1,2}(\alpha+\Delta_{1,2})} = -2|\eta_{1,2} \Omega^\mr{DAC}_{1,2}|^2 \frac{\alpha}{\Delta_{1,2}(\alpha+\Delta_{1,2})}.
\label{eq:qubitshift}
\end{equation}
In the regime where $\alpha \ll \Delta_{1,2}$, Eq. (\ref{eq:qubitshift}) reduces to ${\delta_q^{ss}}$ in Eq. (\ref{eq.Hrwa1}). 
\begin{figure}[tbp]
\centerline{\includegraphics[width=0.5\textwidth]{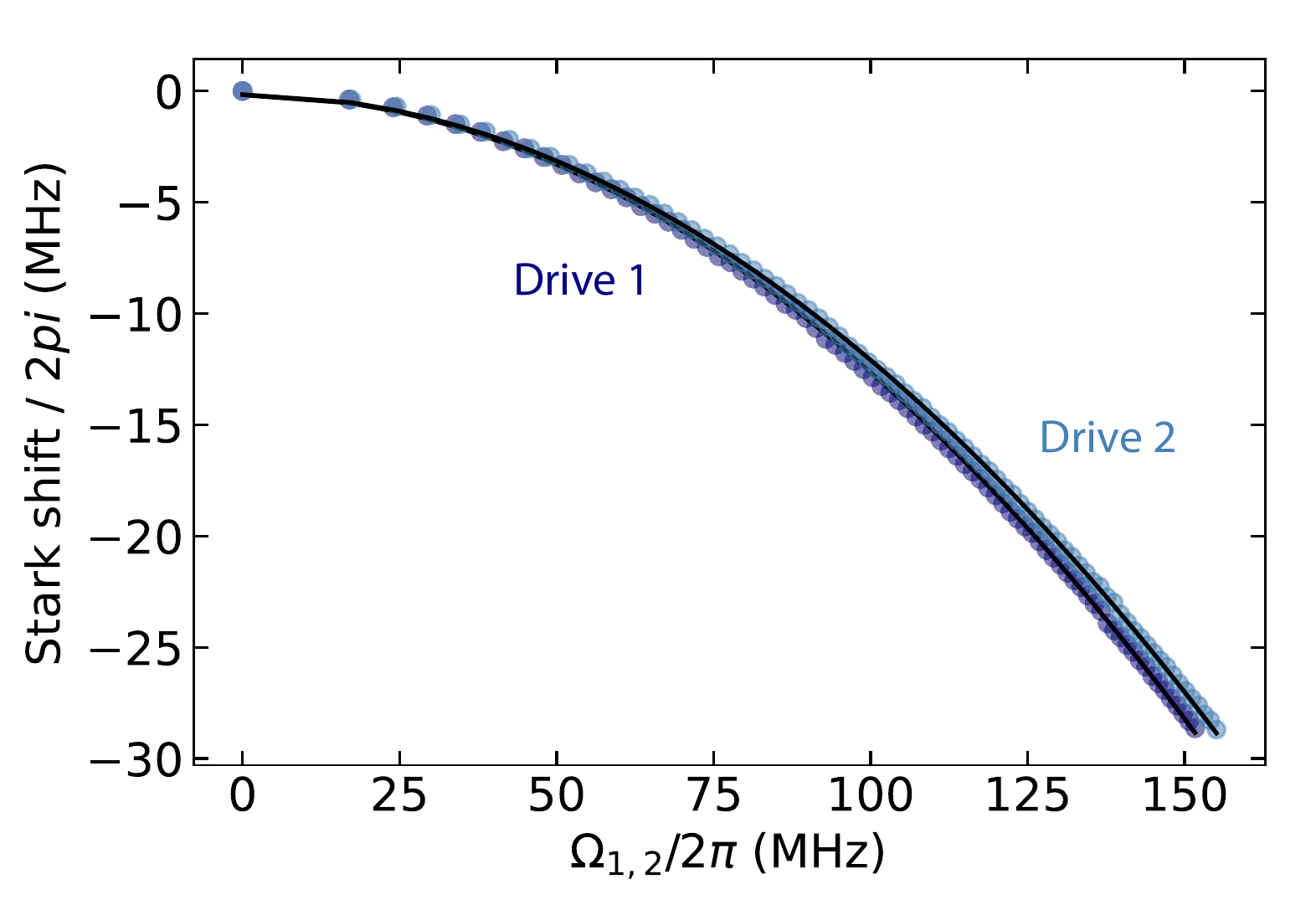}}
\caption{\textsf{\textbf{Stark shift versus drive amplitude $\Omega$.} The extracted qubit frequency is plotted versus drive amplitude for both drive 1 (dark blue) and drive 2 (light blue). Note that in this measurement and throughout the work presented in this paper the drive frequencies are $\omega_1 = \omega_\mr{q} + 2\pi\cdot 492.552\,$MHz and $\omega_2 = \omega_\mr{q} + 2\pi\cdot 505.182\,$MHz, respectively. The fit curves performed with Eq.~(\ref{eq:qubitshift}) are plotted as black lines.}}
\label{fig:SM_Fig1}
\end{figure}

In Supplementary Fig.~\ref{fig:SM_Fig1}, we show the extracted Stark shifts versus drive amplitudes as well as the fit to Eq. (\ref{eq:qubitshift}). 
Note that since the qubit anharmonicty is already known from independent measurements, the conversion factor is the only fitting parameter. 
From the fits we extract $\eta_1 = 2\pi\cdot 256$\,MHz and $\eta_2 = 2\pi\cdot 262$\,MHz. 
These will be used later on to estimate the relative drive amplitudes $\xi_{1,2}$ when the qubit is bichromatically driven.
In the experiments presented in the main text, we use $\Omega^\mr{DAC}_{1}=\Omega^\mr{DAC}_{2}$, resulting in $xi_1\approx xi_2$. 

\section{State tomography}
\label{sec:tomography}
\begin{figure}[b!]
\centerline{\includegraphics[trim = {0cm, 0.5cm, 0.0cm, 0.5cm}, clip=True,scale=0.8]{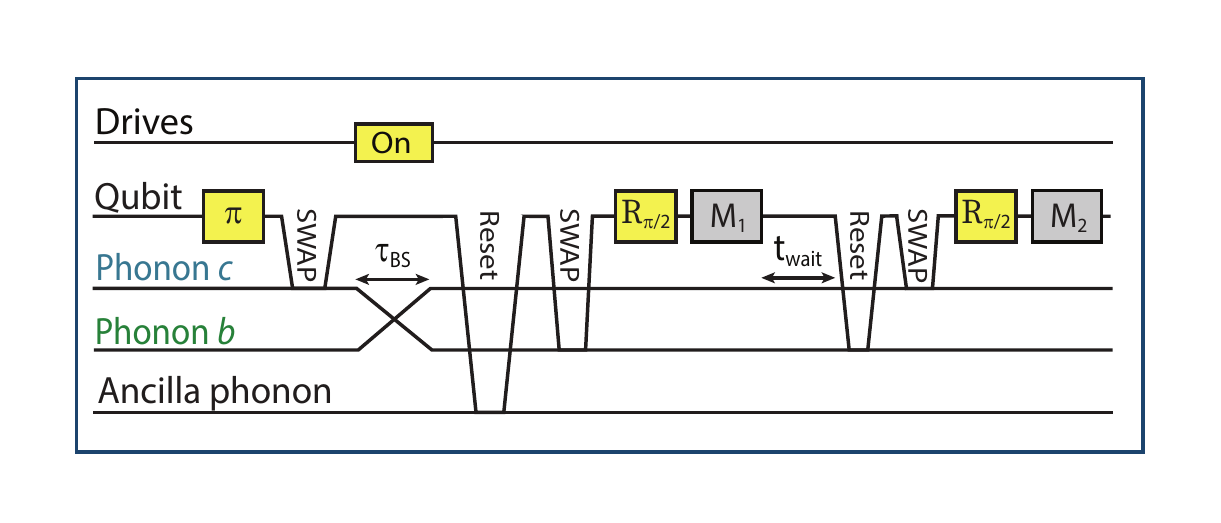}}
\caption{\textsf{\textbf{Beam-splitter and tomography sequence. } We start by loading an excitation into phonon mode $c$. Then, we activate the beam-splitter interaction for the duration of a $\sqrt{\mr{iSWAP}}$-gate and remove any residual qubit population by performing a SWAP with an ancilla mode. Afterwards we swap the state of phonon mode $b$ into the qubit, perform a $\pi/2$-rotation and measure the qubit state. After this operation we wait for $t_\mr{wait}=6\,\mu$s and reset the qubit with the already emptied phonon mode $b$. Finally, we swap the state of phonon $c$ into the qubit, perform another $\pi/2$-rotation, and measure the qubit state again.}}
\label{fig:SM_Fig2}
\end{figure}

In this section, we outline the measurements and data analysis performed for the qubit state tomography shown in Fig.~2\textbf{e} of the main text. 
Following the process described in Ref.~\cite{kjaergaard2020programming}, we measure the different quadratures of the state in the two phonon modes and reconstruct a physical density matrix. As this procedure is well established in the field of cQED, we focus mostly on the determination of our measurement fidelity and cover the rest only in broad strokes. At the end we compute the degree of entanglement using various entanglement metrics. \\

To treat our two phonon modes like two-level systems, we rely on the assumption that the state is mostly contained within their two lowest energy levels. Note that we take into account any leakage into higher Fock levels in the final fidelity error. Under this assumption and in order to measure the joint operators of the two-phonon state after a certain sequence, we repeat the following for each phonon mode: 

\begin{enumerate}
    \item Swap the state of interest from the phonon mode to the qubit via a SWAP operation
    \item Perform a rotation to measure a desired quadrature ($\mathcal{R}^{-\pi/2}_Y$ for $X$, $\mathcal{R}^{\pi/2}_X$ for $Y$, and $\mathcal{I}$ for $Z$) 
    \item Dispersive readout of the qubit through a microwave resonator
\end{enumerate}
 
Between two qubit measurements in each sequence, we wait for a time $t_\mr{wait}=6\,\mu s$ to allow the readout resonator to return to its ground state.
The full sequence, including the state preparation and beam-splitter interaction, is shown in Supplementary Fig.~\ref{fig:SM_Fig2}. 
Together, the two subsequent measurements allow us to measure all two-qubit operators $AB$, where $A,B\in\{I,X,Y,Z\}$. 
We perform 5000 single-shot measurements for each measurement operator, and their average yields the measured probabilities $\mathbf{P_M}=\left(P_{AB}^{00}, ~P_{AB}^{01}, ~P_{AB}^{10},~P_{AB}^{11}\right)$. 
We then convert the measured probabilities into expectation values $\mathbf{m_M}=\left(\langle II\rangle, ~\langle IB\rangle, ~\langle AI\rangle,~\langle AB\rangle\right)$ by using the inverse fidelity matrix $[\beta_a\beta_b]^{-1}$ as follows
\begin{equation}
    \mathbf{m_M} = [\beta_a\beta_b]^{-1} \mathbf{P_M}. 
\end{equation}
By including the fidelities of the different steps of our measurement routine into the fidelity matrices $\beta_{1,2}$, we can exclude the associated infidelities from our measured density matrix $\rho_M$. 
The measured density matrix is then calculated from the measurement operators $A B$ and the corrected expectation values $\langle AB\rangle$ via 
\begin{equation}
    \rho_M = \sum_{A,B} \frac{1}{4}\langle A B\rangle A B. 
\end{equation}
 
To determine the single phonon fidelity matrices $\beta_{a,b}$, we independently measure the fidelities associated with the phonon-qubit SWAP operations, the waiting time after the first measurement, and the qubit state assignment. Table \ref{tab:fidelities} shows the independent measurement used for calibration of each step, as well as the fidelities associated with starting in a $\ket{g}$ or an $\ket{e}$ state in the phonon. With this, we arrive at g[e]-state measurement fidelities of 93.4(2)\% [65.6(6)\%] for phonon $a$ and 93.4(2)\% [69.5(4)\%] for phonon $b$. \\

We emphasize that accounting for the infidelities of our measurement in this way does not guarantee a physical density matrix. 
Therefore, we use a Maximum Likelihood Estimation (MLE) to fit the parameters of the Cholesky decomposition of a physical density matrix $\rho_T$ to the measured density matrix $\rho_M$. This allows to arrive at the density matrix $\rho_\textrm{MLE}$ presented in Fig.~2\textbf{e} of the main paper   \cite{kjaergaard2020programming}. 
\begin{table}[tbp]
\begin{tabular}{l|l|l|l}
\hline
\textbf{Step}          & \textbf{Calibration measurement} & $\ket{g}$-\textbf{Fidelity} & $\ket{e}$-\textbf{fidelity}   \\ \hline
SWAP phonon $a$-qubit      & Vacuum Rabi oscillation between qubit and phonon $a$  & 100\%              & 90.4(3)\%    \\ 
SWAP phonon $b$-qubit      & Vacuum Rabi oscillation between qubit and phonon $b$ & 100\%              & 88.3(3)\%   \\ 
6us wait time & $T_1$-measurement phonon $a$& 100\%              & 92.1(2)\%   \\ 
Qubit state assignment & Single-shot calibration via qubit amplitude Rabi & 93.4(2)\% & 78.8(2)\%   \\ \hline
\end{tabular}
\caption{\textsf{\textbf{Fidelities associated with individual measurement steps. } We calibrate the fidelities of the SWAP-interaction between a phonon mode and the qubit by exciting the qubit and moving it on resonance with the phonon mode. The decaying oscillation lets us fit for the fidelity per SWAP. The fidelity associated with having phonon $c$ idle while we are measuring phonon $b$ and waiting for the cavity to thermalize is given by the decay time of phonon $c$ and the wait time. We quantify the single-shot fidelity of the qubit readout from the contrast of an amplitude Rabi measurement.}  }\label{tab:fidelities}
\end{table}
Finally, to quantify the degree of entanglement of the state we created, we compute the overlap of our reconstructed state with an anti-symmetric Bell state, $\mathcal{F}_\mr{Bell} = \Tr{\sqrt{\sqrt{\rho _T}\rho _\mr{Bell}\sqrt{\rho _T}}}^2=0.69\pm 0.01$. 
For the error on on the overlap, we compute the sampling error by Monte-Carlo propagation of the sampling errors on the measured probabilities $\mathbf{P_M}$, using 5000 sets of randomly probabilities from within $\mathbf{P_M}\pm \sigma_m/\sqrt{n}$, where $\sigma_m$ is the standard deviation of the projective measurements and $n=5000$ is the number of shots per projection operator. 
We then propagate the errors of the infidelity shown in Table \ref{tab:fidelities} and the sampling error to reach the errors given above. 

\section{Three mode interaction parameters}\label{sec:doubleChevron}
In this section we show how we extract the effective coupling strengths and phonon frequency shifts from the phonon population data, an example of which is shown in Fig.~3\textbf{b} and \textbf{c} of the main text. 
We perform a 2D fitting routine based on the following model.
\subsection{Theoretical model}
\noindent The model we use in our fitting routine is based on the equations of motion (EOMs) of five coupled harmonic oscillators. 
We take into account the beam-splitter coupling between all mode pairs as well as their individual frequency shifts and decay rates. 
We keep the naming convention for the three modes presented in the main text, $a$, $b$, and $c$, and consider one additional mode on each side, i.e. we consider, from low to high frequency, the modes $d$, $a$, $b$, $c$, and $e$. 
We assume here that residual direct JC interactions with the qubit are off-resonant and can be neglected. In practice this is only approximately true, but smoothing the data removes the resulting fast oscillating, low amplitude effects and leaves only those relevant to the phonon-phonon interaction. 
To avoid long simulation times when numerically solving the EOMs, we set the frequency of mode $d$ to $\omega_d=0$. 
In addition, we move to a rotating frame in which the effective beam-splitter couplings are time-independent, i.e. $U_\mr{RF}=\exp{-i(\Delta a^\dag a + 2\Delta b^\dag b + 3\Delta c^\dag c + 4\Delta e^\dag e )t}$, where for this section $\D\equiv\mr{FSR}$. 
We vary $\D_{21}-\Delta\equiv \delta$ in our experiment to probe the resonance condition of different phonon-phonon beam-splitter interactions. 
In addition, each phonon experiences a frequency shift $\delta_m$ due to the normal mode splitting with the qubit (main text Eq.~(3)). 
Therefore, the Hamiltonian in the rotating frame is 
\begin{align}
    H_\mr{RF}&=(\delta_a-\delta_d +\delta) a^\dag a + (\delta_b-\delta_d +2\delta) b^\dag b + (\delta_c-\delta_d +3\delta) c^\dag c + (\delta_e-\delta_d +4\delta) e^\dag e \nonumber \\
    & + \sum_{m,k\in\{d,a,b,c,d\}} g_{mk} m^\dag k +\mr{h.c.}\label{eq:eomH}
\end{align}
The first line of Eq. (\ref{eq:eomH}) contains the mode frequency terms and the second line contains the coupling terms between all five modes with effective beam-splitter couplings $g_{mk}$. Here, we consider $g_{mk}\in \mathcal{R}$ following  Eq. (\ref{eq:HJC}). 
Writing $\delta_k-\delta_m\equiv\delta_{mk}$, the EOMs are given by
\begin{gather}
 \begin{pmatrix} \dot{d}(t) \\ \dot{a}(t) \\ \dot{b}(t) \\ \dot{c}(t) \\ \dot{e}(t)\end{pmatrix}
 =-i
  \begin{pmatrix}
   -i\Gamma_d & g_{da} & g_{db} & g_{dc} & g_{de} \\ 
   g_{da} & \delta_{da}+\delta-i\Gamma_a & g_{ab} & g_{ac} & g_{ae} \\ 
   g_{db} & g_{ab} & \delta_{db}+2\delta-i\Gamma_b & g_{bc} & g_{be} \\ 
   g_{dc} & g_{ac} & g_{cb} & \delta_{dc}+3\delta-i\Gamma_c & g_{ce} \\ 
   g_{de} & g_{ae} & g_{eb} & g_{ec} & \delta_{de}+4\delta-i\Gamma_e \\ 
   \end{pmatrix} 
   \begin{pmatrix} d(t) \\ a(t) \\ b(t) \\ c(t) \\ e(t)\end{pmatrix}~, \label{eq:EOMs}
\end{gather}
where $\Gamma_m$ is the decay rate of phonon mode $m$. 
\subsection{Fitting routine}
\begin{figure}[tbp]
\centerline{\includegraphics[trim = {1cm, 0cm, 0.0cm, 0.0cm}, clip=True,scale=0.65]{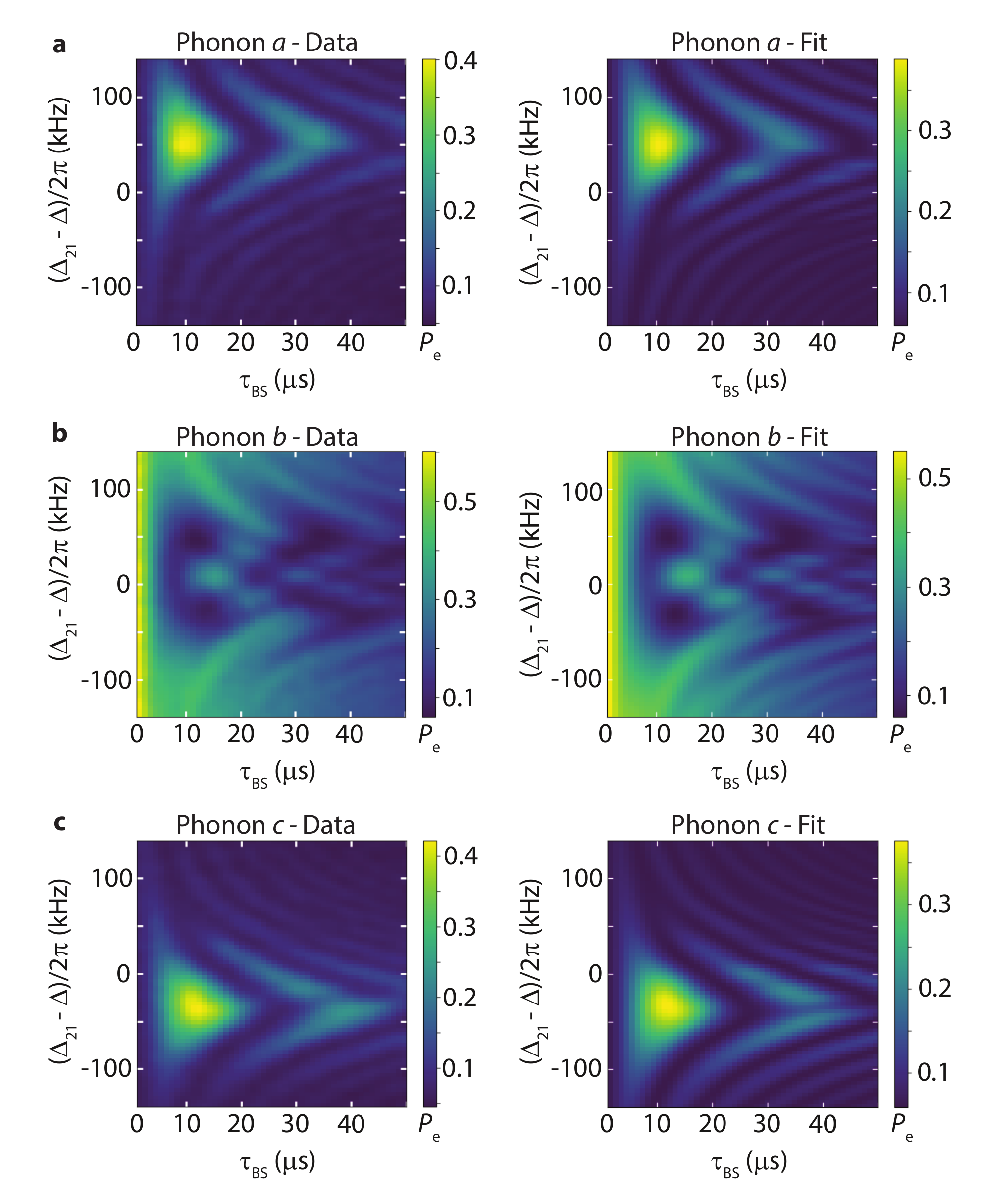}}
\caption{\textsf{\textbf{Example of three-mode coupling data and corresponding fit.} Phonon population measured and fit result for \textbf{a} Phonon mode $a$, \textbf{b} Phonon mode $b$ and \textbf{c} Phonon mode $c$. 
The fit results for this particular dataset are $\delta_{ba}=2\pi\cdot 48\,$kHz, $\delta_{cb}=-2\pi\cdot 31.7\,$kHz, $g_{bc}=2\pi\cdot 17.2\,$kHz, $g_{ab}=2\pi\cdot 20.5\,$kHz, and $g_{ac}=-2\pi\cdot 9.0\,$kHz.}}
\label{fig:SM_Fig3}
\end{figure}

For each value of the modulation depth shown in Fig.~3\textbf{e} of the main text, we measured the phonon populations of the three phonon modes $a$, $b$, and $c$ for 71 equidistant values of $\delta$ between $\pm 2\pi\cdot 140\,$kHz and 100 values of $\tau_\mr{BS}$ up to $50\,\mu$s. 
We numerically solve Eq.~(\ref{eq:EOMs}) and then fit the result to the measured data. Fig.~\ref{fig:SM_Fig3} shows an example of a dataset and the corresponding fit. 
We take into account the infidelity of our SWAP and readout operations by matching the initial phonon populations to the measured populations at $\tau_\mr{BS}=0$ and subtracting the readout infidelity of the qubit $g$-state of $6\%$. 
We fit all four relative phonon shifts $\delta_{md}$ and the couplings between modes $a$, $b$, and $c$ presented in Fig.~3 of the main text, namely $g_{ab}$, $g_{bc}$, and $g_{ac}$. 
To both reduce the fit time and avoid overfitting the data, all other couplings are set equal to the value expected from Eq. (4) of the main text. 
We repeat this for 23 different values of the modulation depth, yielding the data presented in Fig.~3\textbf{e} of the main text. \\

To estimate how resilient our fit is against variation of the resulting parameters, we compute the residuals (sum of squared differences between the solution of the EOMs and the data points) when varying each parameter individually. We then find the value that increases those residuals by 5\% in each direction and use that value as error bar in Fig.~3\textbf{e} of the main text. 

\section{Bright and dark state hybridization}
\label{sec:brightAndDark}
In the main text we present an experiment where three phonon modes are resonantly coupled via effective beam-splitter interactions (main text Fig.~3). 
When measuring the populations in each mode for different interaction times with the three couplings approximately on resonance (i.e. $\Delta_{ab} = \Delta_{bc} = \Delta_{21}$ and $\Delta_{ac}=2\D_{21}$), we observe a typical multi-mode phenomenon, namely the coupling of one mode (in this case mode $b$) to a hybridized state of the two other modes. 
However, if we introduce a small detuning $\delta_\mr{BS}$ to this resonance condition by slightly changing $\Delta_{21}$, we observe an asymmetry between the resonance conditions of the coupling between mode $a$ and $b$ and that of $b$ and $c$. Note that this is a different effect from the phonon frequency shifts described by Eq. (\ref{eq:deltam}). 
In this section we treat the case where the phonon modes are detuned from each other by equal amounts, but where the finite coupling between modes $a$ and $c$ introduces a normal mode splitting between the phonon modes, modifying the resonance conditions. \\  

\begin{figure}[tbp]
\centerline{\includegraphics[trim = {0.5cm, 2cm, 0.0cm, 0.5cm}, clip=True,scale=0.8]{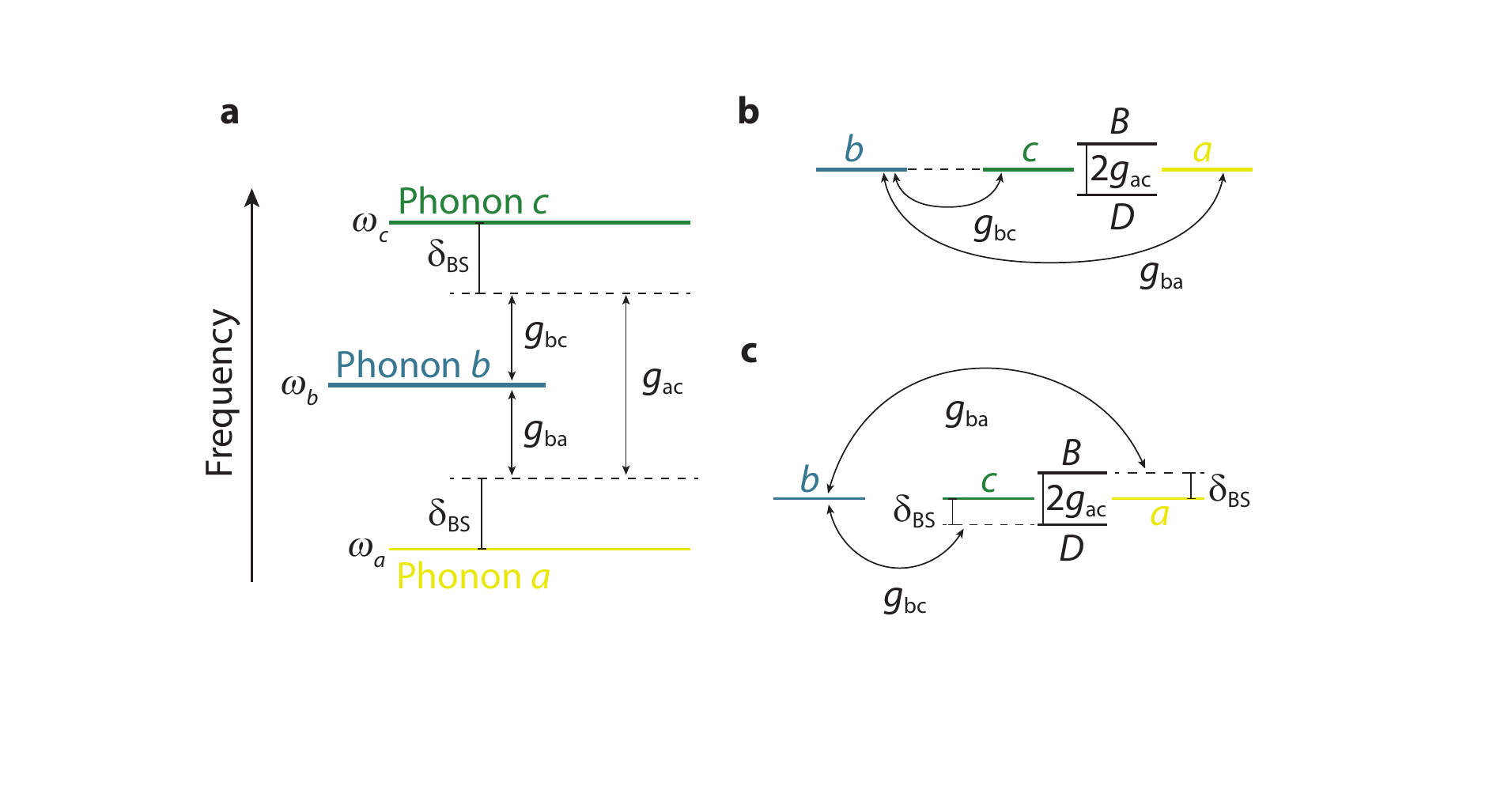}}
\caption{\textsf{\textbf{Energy level diagrams of of three coupled modes.} \textbf{a} Energy diagram in the lab frame with beam-splitter coupling detuned from the resonance conditions by $\dbs$. \textbf{b} The system in the rotating frame of modes $a$, $b$, and $c$, as given by $H_\mr{rotating}$, Eq.~(\ref{eq:Hrotating}) with $\dbs = 0$. and \textbf{c} Same as \textbf{b} with $\dbs = g_{ac}$.   }}
\label{fig:SM_Fig4}
\end{figure}

We start from a system of three harmonic modes $a$, $b$, and $c$ at different frequencies, pairwise coupled via a beam-splitter interaction of the type derived in section \ref{sec:beamsplitterHamiltonian}. 
Mode pairs $(a,b)$ and $(b,c)$ are coupled via neighboring sidebands of the modulated qubit and
mode pair $(a,c)$ is coupled via next-to-neighboring sidebands. 
The beam splitter couplings, in the rotating frame of the three phonon modes, are detuned from their resonance conditions by $\delta_\mr{BS}$ for neighboring sidebands and $2\delta_\mr{BS}$ for next-to-neighboring sidebands, described by the Hamiltonian 
\begin{equation}
H_\mr{rotating} = g_{ab} b^\dag a e^{i\dbs t} + g_{bc}b^\dag c e^{-i\dbs t} + g_{ac} a^\dag c e^{-2i\dbs t} + \mr{h.c.} \label{eq:Hrotating}
\end{equation}
The energy level diagram of this three-level system is depicted in the lab frame in Fig.~\ref{fig:SM_Fig4}\textbf{a} and in the rotating frame of the three phonon modes in Fig. \ref{fig:SM_Fig4}\textbf{b} and \textbf{c}.
The close to resonant coupling $g_{ac}$ leads to a hybridization of modes $a$ and $c$, with the hybridized modes $B=(ce^{-i\dbs t} + ae^{i\dbs t} )/\sqrt{2}$ and $D=(ce^{-i\dbs t} - ae^{i\dbs}t )/\sqrt{2}$. 
Here, $B$ is the hybridized mode arising from the coupling between $c$ and $a$, which we refer to as the bright mode since it couples to mode $b$. On the other hand, $D$ is referred to as dark mode as $b$ does not directly couple to it. 
Furthermore, in our experiment, $g_{ac} \approx g_{bc}\equiv g_\mr{BS}$. 
Thus, we can rewrite the Hamiltonian using the hybridized modes, leading to
\begin{equation}
H_\mr{rotating} = \sqrt{2} g_\mr{BS} (b^\dag B + b B^\dag) + g_{ac} (B^\dag B - D^\dag D)~. \label{eq:Hrotating_brightDark}
\end{equation}
Here we can see that if modes $c$ and $a$ were not coupled, mode $b$ resonantly couples to a superposition of the two. In our case, however, $g_{ac}\neq 0$ and the second term in Eq.~(\ref{eq:Hrotating_brightDark})  does play a role. 
Noting that this second term is diagonal in the basis of $B$ and $D$, we can unveil its effects by entering the rotating frame of the bright and dark modes via the transformation $U_\mr{BD}=\exp\left[ - i g_{ac} (B^\dag B - D^\dag D) t \right]$. In this frame the Hamiltonian is approximately transformed to 
\begin{equation}
H_\mr{BD} = \sqrt{2} g_\mr{BS} (b^\dag B e^{-i g_{ac} t} + b B^\dag e^{i g_{ac} t}) = g_\mr{BS} b^\dag \left(c e^{-i(\dbs+g_{ac})t } + a e^{i (\dbs - g_{ba}) t } \right) + \mr{h.c.} 
\label{eq:BDframe}
\end{equation}
This second rotating frame transformation is exact for $\dbs=0$. 
However, when $\dbs\neq 0$, $B$ and $D$ include a time dependence leading to higher order terms in $H_\mr{BD}$, which we neglect here. 
From the first part of Eq.~(\ref{eq:BDframe}) we can see the behavior for $\dbs=0$, namely a coupling between $b$ and $B$, which is detuned by $g_{ac}$, as drawn in Fig. \ref{fig:SM_Fig4}\textbf{b}. 
In our data, this effect is visible through the reduced contrast of the Rabi oscillations of phonon mode $b$ compared to modes $a$ and $c$, (cf. main text Fig.~3\textbf{d}). More specifically, this shows how an excitation flows from mode $b$ to both $a$ and $c$ and then back to $b$.
From the second part of Eq.~(\ref{eq:BDframe}) it becomes clear that the coupling between modes $b$ and $c$ has a different resonance condition than that between modes $b$ and $a$, which explains the asymmetry with respect to $\dbs$ (or equivalently $\Delta_{21} - \Delta$ as labelled in Fig.~3\textbf{b} and \textbf{c} of the main text). 
We emphasize that this effect arises purely from the non-negligible coupling $g_{ac}$, and is illustrated in Fig. \ref{fig:SM_Fig4}\textbf{c} for the case of $\delta_\mr{BS} = g_\mr{ac}$. 

\end{document}